\documentclass[aip,apl,preprint,floatfix]{revtex4-1}
\usepackage{epsfig,graphicx,pstricks}
\usepackage{svg}
\usepackage{moreverb}
\usepackage{epsfig}
\usepackage{amsmath}
\usepackage{hyperref}
\usepackage[detect-all]{siunitx}
\let\vaccent=\v 
\renewcommand{\v}[1]{\ensuremath{\mathbf{#1}}} 
\newcommand{\gv}[1]{\ensuremath{\mbox{\boldmath$ #1 $}}} 
\newcommand{\abs}[1]{\left| #1 \right|} 
\newcommand{\avg}[1]{\left< #1 \right>} 
\newcommand{\pd}[2]{\frac{\partial #1}{\partial #2}} 
 
\newcommand{\grad}[1]{\gv{\nabla} #1} 
\renewcommand{\div}[1]{\gv{\nabla} \cdot #1} 
\newcommand{\curl}[1]{\gv{\nabla} \times #1} 

\newcommand{\pp}[1]{\left(#1\right)}
\newcommand{\bb}[1]{\left[#1\right]}

\begin{document}

\title{Near-field coupling of gold plasmonic antennas for sub-\SI{100}{nm} magneto-thermal microscopy}

\author{Jonathan C. Karsch}
\affiliation{School of Applied and Engineering Physics, Cornell University, Ithaca, NY, 14853}
\author{Jason M. Bartell}
\affiliation{School of Applied and Engineering Physics, Cornell University, Ithaca, NY, 14853}
\author{Gregory D. Fuchs}
\affiliation{School of Applied and Engineering Physics, Cornell University, Ithaca, NY, 14853}

\begin{abstract}
The development of spintronic technology with increasingly dense, high-speed, and complex devices will be accelerated by accessible microscopy techniques capable of probing magnetic phenomena on picosecond time scales and at deeply sub-micron length scales. A recently developed time-resolved magneto-thermal microscope provides a path towards this goal if it is augmented with a picosecond, nanoscale heat source. We theoretically study adiabatic nanofocusing and near-field heat induction using conical gold plasmonic antennas to generate sub-\SI{100}{nm} thermal gradients for time-resolved magneto-thermal imaging. Finite element calculations of antenna-sample interactions reveal focused electromagnetic loss profiles that are either peaked directly under the antenna or are annular, depending on the sample's conductivity, the antenna's apex radius, and the tip-sample separation. We find that the thermal gradient is confined to \SIrange{40}{60}{nm} full width at half maximum for realistic ranges of sample conductivity and apex radius. To mitigate this variation, which is undesirable for microscopy, we investigate the use of a platinum capping layer on top of the sample as a thermal transduction layer to produce heat uniformly across different sample materials. After determining the optimal capping layer thickness, we simulate the evolution of the thermal gradient in the underlying sample layer, and find that the temporal width is below \SI{10}{ps}. These results lay a theoretical foundation for nanoscale, time-resolved magneto-thermal imaging.
\end{abstract}

\maketitle

\section{Introduction\label{intro}}
Spin-based electronics and high-density magnetic storage require precise control of local magnetic moments in devices\cite{Wolf2001,Challener2009}, often using either applied magnetic fields \cite{beach2005} or spin-transfer torques\cite{yamaguchi2004,mangin2006,madami2011,katine2000}. Development of these technologies will be aided by microscopy techniques, enabling researchers to characterize dynamical, nanoscale magnetic phenomena \cite{bartell2015,freeman2001}, with relevant length scales that are typically \SIrange{10}{200}{nm} \cite{hubert1998} and relevant time scales that are typically \SIrange{5}{50}{ps} \cite{trunk2001,allenspach1990,brien2009}. One existing approach is x-ray magnetic circular dichroism-based microscopy, which offers the desired resolution with spot sizes down to \SI{30}{nm} \cite{acremann2006}. However, it requires a synchrotron facility\cite{stohr1995} and thus it cannot be used in a normal laboratory setting. Another approach is magneto-optical Kerr effect (MOKE) microscopy, which allows for table-top, stroboscopic imaging of spin dynamics with straightforward interpretation\cite{oppeneer1992}. However, the visible to near-IR light that is typically used fundamentally limits the spatial resolution of MOKE to hundreds of nanometers, set by the diffraction-limited focal resolution of approximately half the wavelength \cite{hecht2002}.

We have recently demonstrated a new form of spatiotemporal magnetic microscopy using the time-resolved anomalous Nernst effect (TRANE)\cite{bartell2015,Guo2015,Bartell2017}. In this technique, the anomalous Nernst voltage, $V_{ANE}\propto\abs{\grad T\times\v M}$, is sensitive to a local value of magnetization $\v M$ through a confined thermal gradient $\grad T$\cite{Weiler2012,VonBieren2013}. One benefit of TRANE microscopy is that its spatial resolution is inherited from its local thermal source, which is not fundamentally limited by far-field diffraction. However, recent TRANE studies have relied on focused laser heating, and thus the spatial resolution was still diffraction limited.

To increase the resolution of magneto-thermal microscopy, we theoretically consider focusing electromagnetic radiation to a sub-\SI{100}{nm} region by exciting surface plasmon polaritons (SPPs) on conical plasmonic antennas\cite{Schmidt2012,Babadjanyan2000,Ropers2007,neacsu2010,Muller2016}. This adiabatic nanofocusing on gold antennas has been studied extensively for tip-enhanced Raman scattering\cite{ren2004,liao1982,klingsporn2014,wang2007,lopes2013,kharintsev2011,downes2006}, where electric field enhancements at the apex enable $\sim1000$x enhancements of Raman signals\cite{roth2005,neacsu2010,Issa2007,Stockman2004}. Furthermore, heating confined to a volume well below the diffraction limit with a plasmonic antenna has been studied in the context of heat-assisted magnetic recording\cite{Challener2009,Kryder2008}. For TRANE microscopy, local charge oscillations excited by SPPs at the antenna apex will induce resistive heating in a metallic sample placed in nanoscale proximity.

In this work we theoretically investigate the viability of scanning gold plasmonic antennas as a method to perform nanoscale TRANE microscopy. By studying the non-local, near-field excitation of charge in a flat surface by a conical antenna, we find that electromagnetic loss and the resultant thermal point spread function (PSF) are more nuanced than for focused laser heating. Through finite element calculations of the tip-sample coupling, we reveal that in the range of realistic sample resistivities, apex radii, tip-sample separation, and film thickness, electromagnetic loss is either peaked directly under the apex with a full width at half maximum (FWHM) on the order of the apex radius, or it takes on an annular profile that peaks in a ring $>$\SI{10}{nm} from the center. The potentially large variability in the magnitude and profile of thermal gradients generated in different samples is undesirable for a thermally-based microscopy technique. Therefore, we also perform heat diffusion calculations in which a \SI{10}{nm} platinum capping layer is used as an intermediate heater. We find that centrally-peaked thermal gradients are radially confined to below \SI{100}{nm}, with temporal FWHM below \SI{10}{ps}.

\section{Model background\label{background}}
Our computational model is designed to be representative of an experimental implementation of conical plasmonic antennas in a scan probe microscope (SPM). The antennas are fabricated with an electrochemical etching procedure in hydrochloric acid\cite{kharintsev2011,lopes2013,ren2004}. With appropriate etching voltages and durations, we can experimentally produce conical antennas from \SI{250}{\micro\metre} diameter annealed gold wire (Alfa Aesar), with apex radii as small as \SI{10}{nm} and opening angles near the apex of $\sim$\ang{6}. We use this opening angle in our calculations, and consider tip radii $>$\SI{10}{nm}. Tip-sample separations (gap values) are based on the SPM feedback mechanism, which leads to distances of closest approach of order \SI{1}{nm}. More details on the computational model are in Supplementary Information Section S.1.

Experimentally, far-field light is coupled into SPPs at the surface of the antenna with diffraction gratings of period
\begin{equation}
\label{gratea}
a=\bb{\pp{\frac{\epsilon_r'}{\epsilon_r'+1}}^{1/2}-\sin\theta}^{-1},
\end{equation}
where $\epsilon_r'$ is the real part of the dielectric constant of gold and $\theta$ is the angle of incidence\cite{Raether1988}. Numerical simulations suggest that light is best coupled into SPPs for a sinusoidal grating\cite{Hessel1965} of depth \SI{180}{nm}; we use a grating of this shape in our model to excite SPPs at the antenna boundary with \SI{780}{nm} light.

\begin{figure}[b]
\[\includegraphics[width=13.3cm]{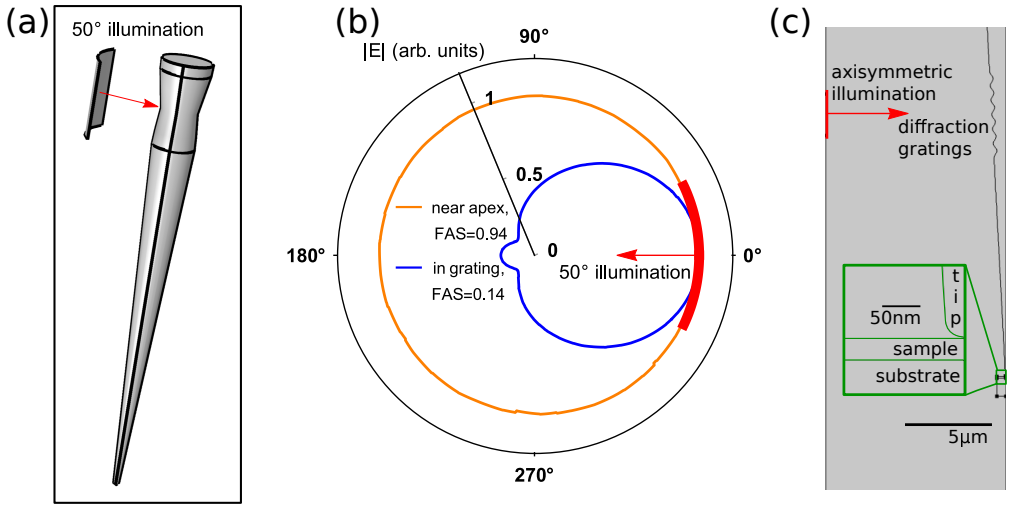}\]
\caption{\label{illumandgeo} \bf Illumination geometry. \rm (a) Far-field light is coupled into the plasmonic modes by asymmetric illumination with a microscope objective of \ang{50}. (b) Electric field distribution from 3D calculations at the grating (blue) and \SI{2}{nm} away from the apex, in a \SI{15}{nm} radius circle (orange). Field asymmetry (FAS) is shown for each. (c) 2D axisymmetric model geometry used for calculations, which captures the antenna taper but ignores asymmetry.}
\end{figure}

In a SPM, grating illumination will not be axially symmetric because the tip is only illuminated from one side. Therefore, we initially simulate the electric field around the apex of a 3D antenna illuminated over $50^\circ$, as in Fig. \ref{illumandgeo}(a). We define the field asymmetry as the minimum value of the electric field magnitude, $\abs{\v E}$, around the antenna divided by $\abs{\v E}$ at $0^\circ$ (center of illumination). We find a value of 0.94 near the apex, in sharp contrast to the initial asymmetry of 0.14 at the grating, shown in Fig. \ref{illumandgeo}(b). This justifies the 2D axisymmetric simulation geometry that we use for subsequent calculations because it is more computationally efficient (Fig. \ref{illumandgeo}(c)).

To calculate the thermal response of the sample to the plasmonic excitations, we consider loss dominated by Ohmic heating, calculated as the product of electric field $\v E$ and current $\v J$ as
\begin{equation}
\label{losseq}
W=\tfrac{1}{2}\pp{J_r^*\cdot E_r+J_z^*\cdot E_z},
\end{equation}
where asterisks denote the complex conjugate. Contributions from dielectric loss, arising from the imaginary part of the dielectric constant, are ignored because, for metallic samples, they are insignificant compared to conductive losses as long as the frequency is away from a resonance. We treat the gold antennas as lossless ($\sigma=0$, $Im[\epsilon]=0$) to decouple effects of the SPP decay length from the size of the model geometry. Effects of finite gold conductivity are treated in Section S.2. Although we perform our calculations with a $6^\circ$ cone angle, SPP focusing and dielectric losses are weakly dependent on cone angle. In Supplementary Sections S.2 and S.3 we justify our approximations and discuss their implications.

\section{Results\label{results}}
\subsection{Heat source point spread function\label{heatPSF}}
\begin{figure}[b]
\[\includegraphics[width=13.3cm]{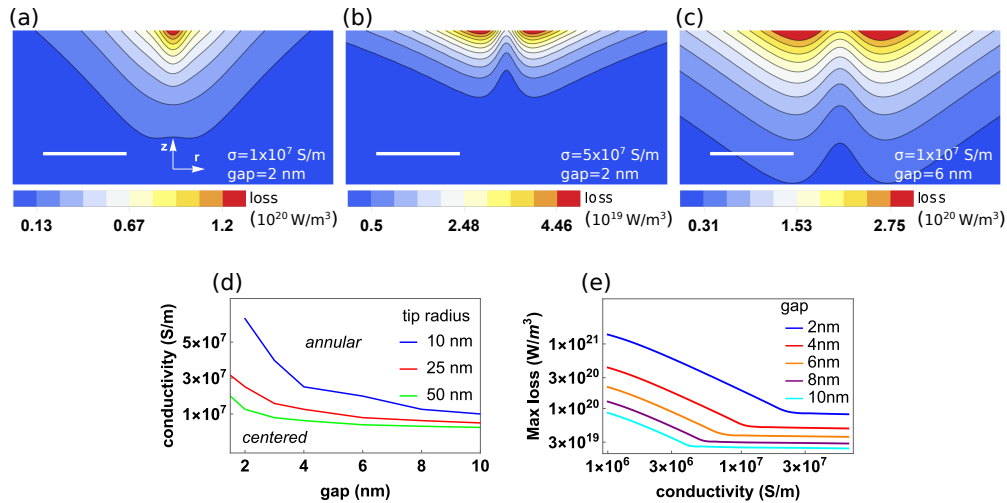}\]
\caption{\label{PSFfig} \bf Loss point spread function. \rm Loss profiles for a \SI{30}{nm} metal sample -- 2D axisymmetric is data reflected across the central axis for visualization. (a) $\sigma=$\SI{1e7}{S/m} and \SI{2}{nm} gap, (b) $\sigma=$\SI{5e7}{S/m} and \SI{2}{nm} gap, and (c) $\sigma=$\SI{1e7}{S/m} and \SI{6}{nm} gap. All horizontal scale bars are \SI{50}{nm}, and all plots are \SI{10}{nm} tall.  (d) Transition boundary between centered and annular profiles, seen in (a)-(c). Values below and to the left of the curves correspond to centered profiles, and those above and to the right correspond to annular profiles. (e) Dependence of loss magnitude on conductivity and gap for a \SI{25}{nm} apex radius.}
\end{figure}

The spatial distribution of the electromagnetic loss determines the thermal gradient in near-field heating. Depending on the tip apex radius, tip-sample distance, and sample conductivity, loss is either centered under the apex or distributed in an annulus. Fig. \ref{PSFfig}(a)-(c) present loss profiles for a characteristic tip with a \SI{25}{nm} radius of curvature above a metallic sample with permittivity $\epsilon_r=-20$ (representative of permalloy\cite{Tikuisis2017}) and varying conductivity. In Supplementary Section S.4 we show there is a very weak relationship between loss profile and sample permittivity, which is insignificant for realistic sample conductivity. These calculations are performed in the frequency domain, with \SI{1}{W} input power coupled into an antenna as discussed in Section \ref{background}. We first consider a \SI{30}{nm} sample grown on a sapphire substrate. This allows us to focus on tip-sample interactions without coupling to the dielectric substrate, which we find can be non-trivial for much thinner samples.

For a realistic tip-sample distance of \SI{2}{nm} -- measured from the center of the tip apex to the top of the sample plane -- and a sample conductivity of \SI{1e7}{S/m} (representative of Pt), the electromagnetic loss has a point spread function (PSF) that has its maximum centered under the tip. A centered loss profile will produce a centered thermal gradient profile. However, increasing either the conductivity or the gap size leads to annular loss. The width of the loss PSF is minimized when the profile is centered, leading to the best spatial resolution. The loss is also more confined to the surface for increased conductivity (Fig. \ref{PSFfig}(b)), while larger gaps noticeably decrease the magnitude of the loss (Fig. \ref{PSFfig}(c)).  Fig. \ref{PSFfig}(d) presents the transition as a phase diagram, where for each apex radius, the curve separates regions of centered and annular profiles depending on the gap size and the sample conductivity.

The magnitude of loss decreases with increasing gap size, but it has a less straightforward dependence on sample conductivity. Fig. \ref{PSFfig}(e) shows peak loss for varying conductivity at different gaps with a \SI{25}{nm} radius tip. Peak loss decreases with increasing conductivity until it reaches the centered-to-annular transition, above which the loss profile becomes more confined to the surface but does not diminish. Maximum power dissipation for the centered profiles is comparable to those simulated for laser heating\cite{bartell2015}.

\begin{figure}[b]
\[\includegraphics[width=13.3cm]{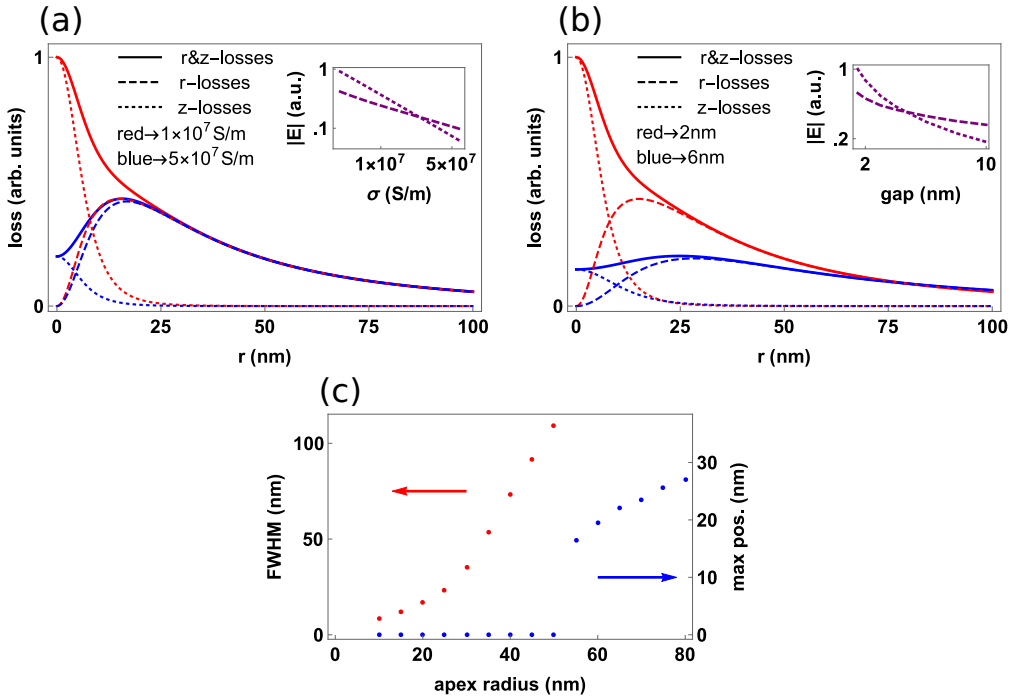}\]
\caption{\label{lossfig}\bf Radial and vertical losses for \SI{25}{nm} apex. \rm When the conductivity (a) or gap (b) increases past a certain value (i.e., crosses the line in Fig. \ref{PSFfig}(d)), the total loss profile (solid lines) transitions from centered (red) to annular (blue). Vertically excited loss is shown with dotted lines, and radially excited loss is shown with dashed lines. (a) and (b), inset, vertical (dotted) and radial (dashed) maximum electric field magnitudes along the sample surface. (c) Peak position and FWHM of loss profiles for increasing tip radius at realistic scan probe gaps. Once the peak is off-center we do not calculate the FWHM.}
\end{figure}

We can understand the profile transition by considering separately the losses excited by vertical and radial electric fields. Fig. \ref{lossfig} shows how the shape of the loss profile depends on the relative magnitude of the individual $r$ and $z$ contributions. We look at total loss and decompose it into vertically and radially excited losses, taken along half-line cuts (moving radially out from directly under the apex) at the sample surface. Curves in Fig. \ref{lossfig}(a) and (b) correspond to representative conductivity (a) or gap (b) values where the profile is centered or annular, according to Fig. \ref{lossfig}(d). Vertically excited loss is always centrally-peaked, whereas radially excited loss is always annular. When the conductivity or gap is increased, the peak of the $r$ loss is larger than that of the $z$ loss, and the profiles become annular (Fig. \ref{lossfig}(a,b)). The $r$ loss is also more extended in the radial direction than $z$ loss, therefore their relative magnitude will non-trivially determine the width of the PSF.

To understand the origin of these loss profiles better, we plot the peak electric field along the same radial line cut for varying conductivity and gap distance (inset in Fig. \ref{lossfig}(a) and (b), respectively). In each case, the point where the vertical component drops below the radial component corresponds to the point in Fig. \ref{PSFfig}(d) where the loss profile transitions from centered to annular. Physically, a larger conductivity leads to larger charge concentration at the sample surface, which screens vertical oscillations more effectively than radial ones. In the case of a large gap, the charge distribution decreases in magnitude and spreads, leading to the generally steeper decrease of $z$ losses seen in Fig. \ref{lossfig}(b) inset.

Next we investigate the effect of apex radius on profile size. We consider a Pt sample with a Au tip of varying radius placed between \SI{1.3}{nm} (for \SI{10}{nm} apex) and \SI{2.6}{nm} (for \SI{80}{nm} apex). We choose to vary the distance with the radius to model the FWHM in a realistic experimental setting, where gap distance depends on the interatomic forces between the apex and the sample\cite{Butt2005}, and is sensitive to apex radius\cite{Hamaker1937}. In Fig. \ref{lossfig}(c) the position of the peak loss remains at zero (centered) until an apex radius of \SI{55}{nm}, where it moves off-center. This is due to a wider charge concentration in larger tips. It is useful to look at the FWHM for centered profiles, shown in Fig. \ref{lossfig}(d), which shows two regimes of increasing width. The FWHM grows with apex radius at \SI{0.95+-0.09}{nm/nm} for radii below \SI{30}{nm}. For radii above \SI{30}{nm}, the FWHM grows as \SI{3.73+-0.03}{nm/nm}.

\begin{figure}[b]
\[\includegraphics[width=13.3cm]{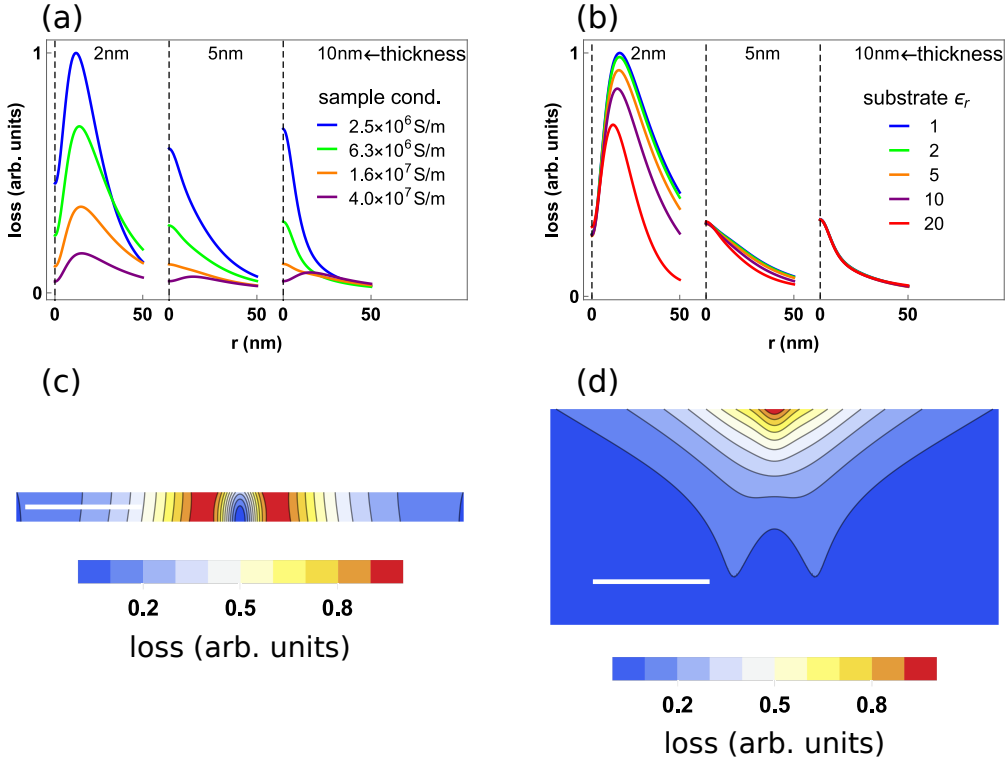}\]
\caption{\label{samplethickness} \bf Sample thickness. \rm (a) Loss profiles at different sample thicknesses for varying sample conductivity. (b) Loss profiles for different sample thicknesses for varying substrate permittivity. (c) Full sample losses for \SI{2}{nm} sample, $\epsilon_r=10$. (d) Full sample losses for \SI{10}{nm} sample, $\epsilon_r=10$. Horizontal scale bars are \SI{50}{nm}, and (c) and (d) are \SI{2}{nm} and \SI{10}{nm} tall, respectively.}
\end{figure}

In samples with thicknesses comparable to the penetration depth of electromagnetic loss, the loss profile is significantly altered. We fix the apex radius at \SI{25}{nm} and the gap at \SI{2}{nm}, then consider a range of realistic sample conductivities at several sample thicknesses. The results are plotted in Fig. \ref{samplethickness}(a). For a \SI{2}{nm} sample, loss is radially annular for all values of conductivity shown (Fig. \ref{samplethickness}(c)). As thickness increases (moving to the right in Fig. \ref{samplethickness}(a)), loss profiles approach the centered distribution shown in Fig. \ref{PSFfig} for the \SI{30}{nm} sample for thicknesses above \SI{10}{nm} (Fig. \ref{samplethickness}(d)). The dielectric constant $\epsilon_r$ of the substrate material also impacts loss profiles for films thinner than \SI{10}{nm} (Fig. \ref{samplethickness}(b)). The large variation in loss profiles for film thicknesses of interest for TRANE microscopy further complicates the uniformity of a plasmonic antenna probe.

\subsection{Heating with a capping layer\label{capheating}}
To reduce the sample-to-sample variability demonstrated above, we propose the use of an intermediate heater layer. We examine platinum because (1) it has conductivity commensurate with centered losses for achievable tip radii, and (2) it is often already present in spin-Hall devices.

\begin{figure}[b]
\[\includegraphics[width=13.3cm]{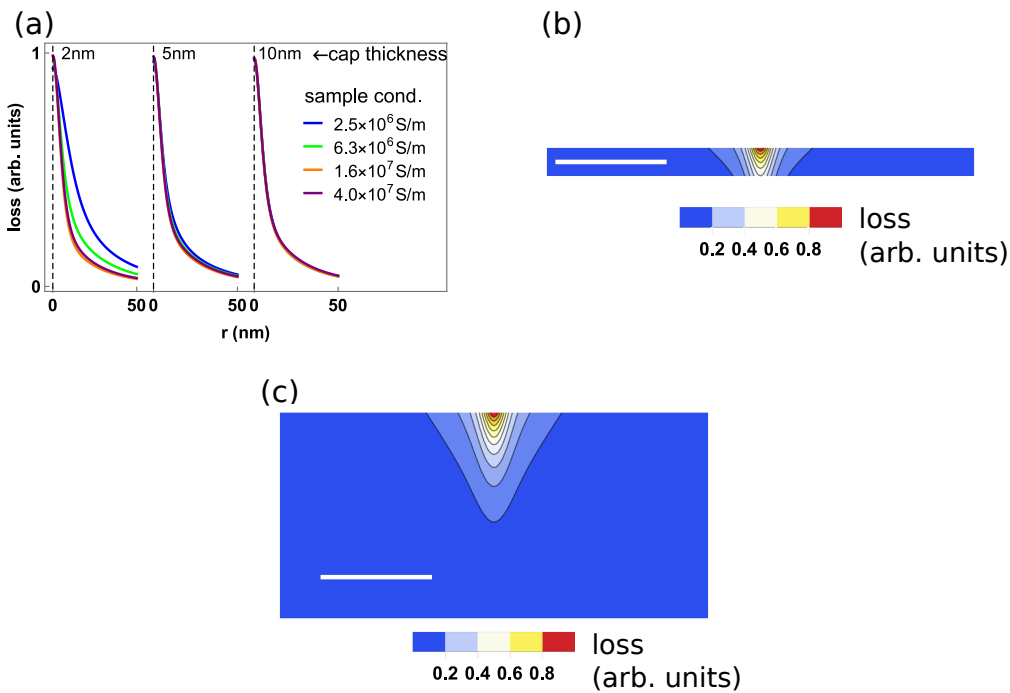}\]
\caption{\label{capthickness}\bf Capping layer thickness. \rm (a) Loss profiles along capping layer surface at selected capping thicknesses for varying sample conductivities. (b) Full capping layer loss in a \SI{2}{nm} thick layer. (c) Full capping layer loss in a \SI{10}{nm} thick layer. Horizontal scale bars are \SI{50}{nm}, and (b) and (c) are \SI{2}{nm} and \SI{10}{nm} tall, respectively.}
\end{figure}

We first determine an optimal capping size that is thick enough to decouple the Pt losses from the material underneath, but thin enough that most of the thermal energy is transfered to the sample. By varying sample conductivity for various cap thicknesses, we find a \SI{10}{nm} layer is the thinnest for which there is negligible effect from sample conductivity (Fig. \ref{capthickness}(a)). By examining loss profiles at the top of the Pt cap for a constant sample conductivity (lines of the same color), we reveal that sample thickness variations below \SI{10}{nm} affects the loss confinement, most notably at low conductivity. The peak of loss shows slight dependence on conductivity for thin samples, but a trivial dependence on sample thickness. In fig. \ref{capthickness}(b) we plot the 2D loss profile for a \SI{2}{nm} sample; loss is still significant at the bottom of the capping layer, so the underlying sample influences the profile. This is in contrast to the case shown in Fig. \ref{capthickness}(c), where the loss in a \SI{10}{nm} sample is not affected by the sample.  Above \SI{10}{nm} the profiles no longer change with sample thickness. Furthermore, line cuts with varying conductivity converge to the same centered loss profile, suggesting that a \SI{10}{nm} thick capping layer can provide the same heat source to a variety of samples of interest. This is significant for developing a reliable microscopy technique. 

To investigate thermal gradient generation, we simulate loss in the Pt capping layer when an antenna (\SI{15}{nm} apex radius) is excited with a Gaussian pulse of \SI{3}{ps} FWHM centered at \SI{9}{ps} in the calculation. We ignore direct heat transfer between the gold tip and Pt layer. While electromagnetic loss may heat the gold tip in excess of \SI{100}{K}, the thermal conductivity of air -- $\sim$\SI{1e-8}{W/m.K} at the nanometer scale\cite{Zhang2009} -- is low enough such that there is negligible heat transfer across the air gap compared to the heat deposited through electromagnetic loss in the Pt. Near-field heat transfer between Au and Pt for similar geometries has also been shown to be orders of magnitude smaller than our simulated Pt losses\cite{Kittel2005}.

\begin{figure}[h]
\[\includegraphics[width=13.3cm]{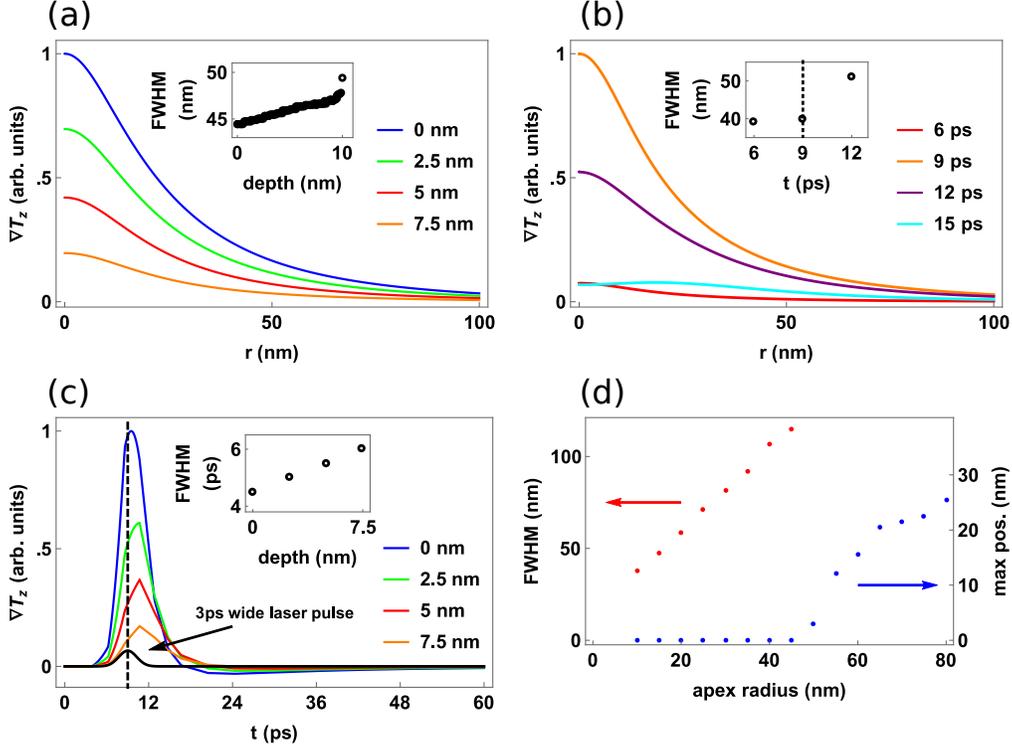}\]
\caption{\label{gradientsfwhm} \bf Thermal gradients with capping layer. \rm For a tip apex of \SI{15}{nm}, the Pt capping layer heats up an underlying permalloy sample. (a) Thermal gradients at selected depths within the sample. Inset: FWHM as a function of depth. (b) The gradient at the sample surface for various times around peak laser power (\SI{9}{ps}), FWHM for centrally-peaked profiles inset. (c) Temporal evolution of $\grad T_z$ at selected depths directly under the apex, with temporal widths remaining within \SI{6}{ps} (inset). (d) Peak position and FWHM of $\grad T_z$ profiles for increasing tip radius at realistic scan probe gaps. Once the peak is off-center we do not calculate the FWHM.}
\end{figure}

We now investigate the heating of a characteristic \SI{10}{nm} permalloy sample with a \SI{10}{nm} Pt capping layer. For focused laser heating, we expect a depth-dependent thermal gradient because light is absorbed in the sample on a scale set by its skin depth. Here, the near-field electromagnetic loss within the cap generates thermal gradients in the magnetic layer, thus the expected profile is less obvious. The thermal gradient varies strongly with depth in the sample. In particular we see a more than 50\% decrease in the thermal gradient magnitude halfway through the depth of the \SI{10}{nm} film, and a more than 75\% decrease in the lower quarter, shown in Fig. \ref{gradientsfwhm}(a). The TRANE voltage is proportional to the in-plane magnetic moment; the depth dependence could present complications from non-uniform moment distributions, where the final signal is a weighted average of the varying in-plane moments. However, typical thin-film magnetic materials have uniform magnetization along their thickness direction, an assumption we will apply in our following analysis.

To determine the spatial resolution of scanning TRANE microscopy with a Pt cap, we calculate the width of the vertical thermal gradient, $\grad T_z$ (Fig. \ref{gradientsfwhm}(a), inset). We find that the average FWHM is \SI{45.3}{nm}, weighted by peak gradient value, calculated when the thermal gradient is at its maximum value. As seen in Fig. \ref{gradientsfwhm}(a), inset, the width increases more rapidly as we approach the interface with the substrate. This is likely due to the interfacial thermal resistance, which is approximately \SI{1e8}{K.m^2/W} for a metal-sapphire boundary\cite{2-Wang2007}. However, the thermal gradient magnitude is also diminished as a function of distance from the surface, thus the wider profile at larger depths does not significantly impact the weighted average width.

We also consider the time evolution of $\grad T_z$. In Fig. \ref{gradientsfwhm}(b), $\grad T_z$ is plotted for a selection of times before and after the center of the laser pulse. The spatial width (inset) significantly increases at \SI{12}{ps}, where it is 25\% larger than the FWHM at the peak value of $\grad T_z$. In general, the width increases with time. We do not calculate the width at \SI{15}{ps} because by then the thermal gradient peak is off-center. This temporal broadening of the PSF is offset by the fact that $\grad T_z$ is less than 10\% of its peak value by this time, and thus it will have a small effect on the spatial resolution

The thermal gradient is centrally-peaked for a majority of its duration, so we calculate the temporal width based on the evolution of $\grad T_z$ at different depths along a central vertical line in the sample, in Fig. \ref{gradientsfwhm}(c). For reference, the shape of the laser pulse is presented as a solid black curve, with the peak time denoted with a vertical dashed black line. The thermal gradient peaks around \SI{10.7}{ps}, \SI{1.7}{ps} after the peak of the pulse. The temporal FWHM shows depth variation (inset) of a couple picoseconds. Due to the transfer of heat down through the sample, at a certain time (around \SI{20}{ps}, but it is depth-dependent) the thermal gradient reverses sign. Although the dip is much smaller than the peak, this will reduce the overall TRANE signal depending on how it is measured.

The profiles in Fig. \ref{gradientsfwhm}(a)-(c) were calculated with a \SI{15}{nm} apex radius; we now consider the effects of varying tip radii on the spatial extent of the thermal PSF. Fig. \ref{gradientsfwhm}(d) shows the FWHM of the $\grad T_z$ along the sample top at its temporal peak. For apex radii of \SI{45}{nm} and below, the gradient profile is centered with a FWHM of \SI{14.0+-1.2}{nm}+\num{2.3+-0.1}\,$r_a$, where $r_a$ is the apex radius. There are no separate discernible regions as with the FWHM of the loss profile (Fig. \ref{lossfig}(c)). Due to the radial spread of heat, the thermal width is in all cases wider than the loss width, however it remains confined to a region approximately the size of the apex. For a \SI{90}{nm} diameter apex -- the largest apex size with a centered profile -- the FWHM is \SI{115}{nm}.

\section{Summary}
The nanofocusing of light with conical plasmonic antennas offers a means to access nanoscale electromagnetic heating with far-field excitation, making it a realistic path to improving the spatial resolution of TRANE microscopy to nanoscale lengths. We have shown that, when a sub-\SI{100}{nm} antenna apex is within nanometers of a sample surface, strong near-field coupling produces electromagnetic loss in the sample that is confined to a region comparable to the size of the apex. Additionally, the loss profile is dependent on the sample conductivity and the tip-sample distance. As these parameters increase, vertically excited loss decays more quickly than radially excited loss. Radially excited loss is peaked off-center, and eventually it overtakes vertical loss, forming an annular loss profile. For films with thicknesses similar to or below the penetration depth of loss, the bottom interface can have a profound effect on the loss profile. In the case of a \SI{2}{nm} film on sapphire, the loss profile entirely loses its central peak. This thickness sensitivity persists until the film is thicker than \SI{10}{nm}.

To avoid sample-to-sample variation of resolution and thermal gradient amplitude, we propose the use of a Pt capping layer. We find that \SI{10}{nm} of Pt is sufficient to act as an intermediate heater for the underlying material. Because losses in the capping layer are separated from the electromagnetic properties of the sample, the cap may heat uniformly across different samples. These results point the way toward accessible spatiotemporal magnetic microscopy with sub-\SI{100}{nm} spatial resolution and sub-\SI{10}{ps} temporal resolution -- capabilities that can enable the next breakthrough in energing spintronic technologies.

\section*{Acknowledgments}
This research is supported by the U.S. Air Force Office of Scientific Research under Contract No. FA9550-14-1-0243.

\section*{Supplementary information}
\renewcommand{\thesection}{S.\arabic{section}}
\renewcommand{\thefigure}{S.\arabic{figure}}
\renewcommand{\theequation}{S.\arabic{equation}}
\setcounter{section}{0}
\setcounter{figure}{0}
\setcounter{equation}{0}

\section{Computational model}
We use COMSOL Multiphysics\textsuperscript{\textregistered} with the electromagnetics (emw) and heat transfer (ht) modules. The emw module calculates the electric field $\v E$ by solving Maxwell's equations in the frequency domain as
\begin{equation}
\label{emw}
\curl\mu_r(\curl \v E)-k_0^2\pp{\epsilon_r-\frac{j \sigma}{\omega \epsilon_0}}\v E=0,
\end{equation}
where $\mu_r$ is relative permeability, $k_0$ is free-space wave number, $\epsilon_r$ is relative permittivity, $\sigma$ is conductivity, $\omega$ is angular frequency, and $\epsilon_0$ is permittivity of free space. Electromagnetic loss is then calculated according to Eq. 2. The ht module calculates temperature $T$ according to
\begin{equation}
\label{ht}
D\, C_p\pd{T}{t}-\div (k\grad T)=Q,
\end{equation}
where $D$ is density, $C_p$ is heat capacity at constant pressure, $k$ is thermal conductivity, and $Q$ is the heat source. In our models $Q$ is defined as the electromagnetic loss calculated with the emw module.

The finite element mesh size in the 2D axisymmetric models is at most one-fourth the wavelength ($\lambda$). Through mesh refinement studies, we determine sufficient maximum element sizes along the edges of the antenna and the sample. Based on convergence (within 1\%) of the electric field magnitude at the apex, a maximum size of $\lambda/30$=\SI{26}{nm} is chosen along the antenna and apex edges. Although this is about the size of the apex radius, it is only setting the maximum, and the meshing algorithm produces much smaller elements along the curved apex. Along the sample and capping layers a maximum size of \SI{0.5}{nm} is sufficient, based on convergence of electromagnetic loss along the layer's surface. The inner edge -- in the 2D axisymmetric geometry shown in Fig. 1(c) -- of the gap is set to a maximum of \SI{0.7}{nm} according to the same metric.

\section{\label{goldsigma}Effect of finite conductivity of gold}
In Section III we treat the gold antenna as non-conducting; surface plasmon polaritons (SPPs) propagate on the gold-air boundary, but do not suffer electromagnetic loss in our model. Here we look at changes to the loss point spread function (PSF) from turning on antenna conductivity, as well as apex heating and the resultant thermal expansion.
\subsection{Loss profiles}
\begin{figure}[b]
\[\includegraphics[width=13.3cm]{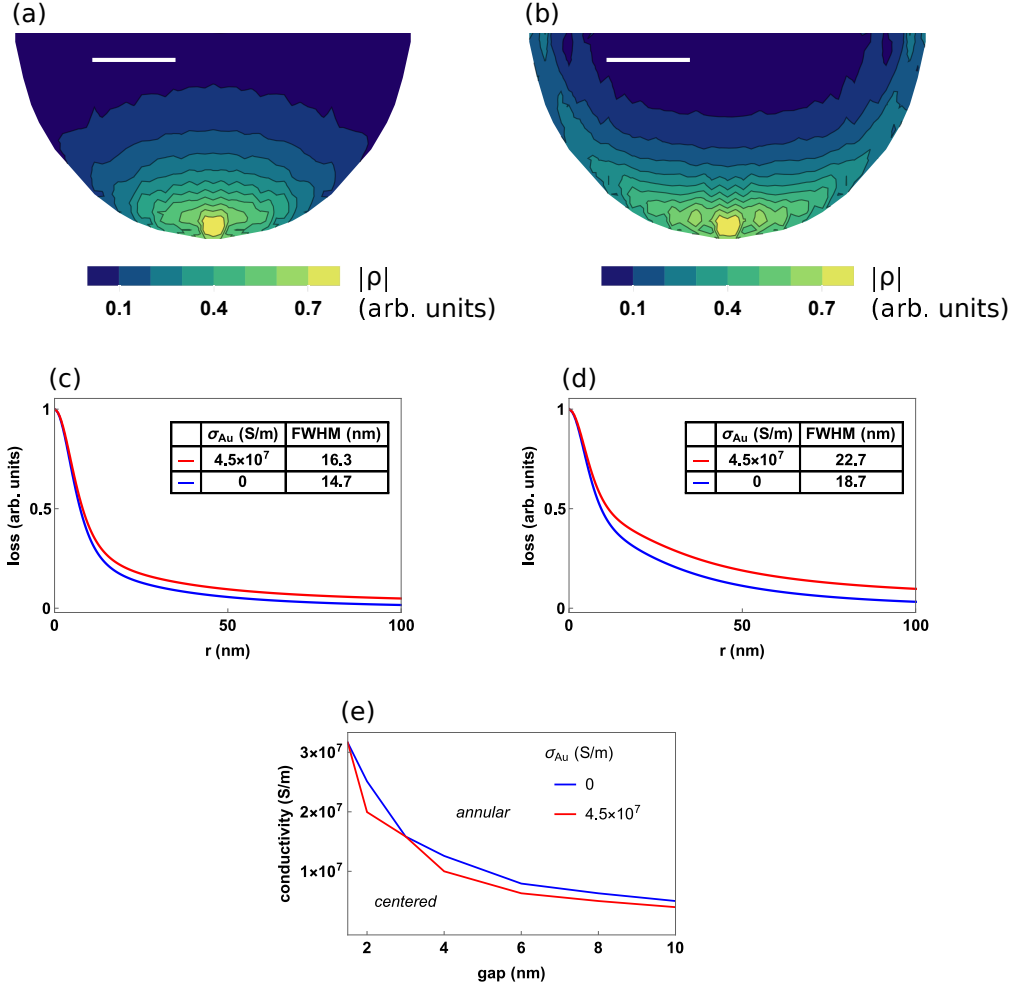}\]
\caption{\label{condgold} \bf Charge and loss with tip conductivity. \rm Charge oscillations in the tip apex are more centrally concentrated in a non-conductive tip (a) than a conductive one ($\sigma_{Au}=$\SI{4.5e7}{S/m}) (b); scale bars are \SI{10}{nm}. (Jagged contours arise from calculating $\abs{\rho}\propto\abs{\div\v J}$, the divergence of current density, on a finite mesh.) This tends to broaden sample loss profiles, in (c) and (d) for sample conductivity of \SI{5e6}{S/m} and \SI{1e7}{S/m}, respectively. All profiles are individually normalized by their maximum value; losses are smaller when accounting for conductivity because of energy dissipation along the tip (see Section \ref{sppprop}). (c) The central-to-annular profile transition for zero (blue) and finite (red) gold conductivity.}
\end{figure}

The loss profile is in part determined by charge distribution in the apex, which we expect is altered by the inclusion of conductivity. Fig. \ref{condgold}(a) and (b) show the charge distribution $\rho$ in a \SI{25}{nm} radius apex for zero and finite gold conductivity ($\sigma_{Au}$=\SI{4.5e7}{S/m}). The finite conductivity pushes charge towards the surface of the apex, shifting the average radial charge position, $\avg{r_{\abs{\rho}}}$, from \SI{7.0}{nm} to \SI{10.1}{nm}. However, the loss profile in the sample qualitatively maintains its shape. In Fig. \ref{condgold}(c) and (d) we plot loss and the full widths at half maximum (FWHM) with zero and finite conductivity for a \SI{25}{nm} apex and a \SI{1.75}{nm} gap, for sample conductivities of \SI{5e6}{S/m} and \SI{1e7}{S/m}. The FWHM is larger with finite conductivity, related to the centrally focused-to-annular transition of the profile. In Fig. \ref{condgold}(e), the transition line for a \SI{25}{nm} apex and finite conductivity is compared to that from Fig. 2(d). We see that for finite conductivity, the profile tends to turn at slightly smaller conductivities and smaller gaps. Thus for the same sample conductivity and gap, the ratio of radially excited loss to vertically excited loss is larger when $\sigma_{Au}\neq0$. This explains why -- for the same sample conductivity -- the profiles in Fig. \ref{condgold}(c) and (d) are wider with $\sigma_{Au}\neq0$.

\subsection{Heating the tip}
\begin{figure}[b]
\[\includegraphics[width=13.3cm]{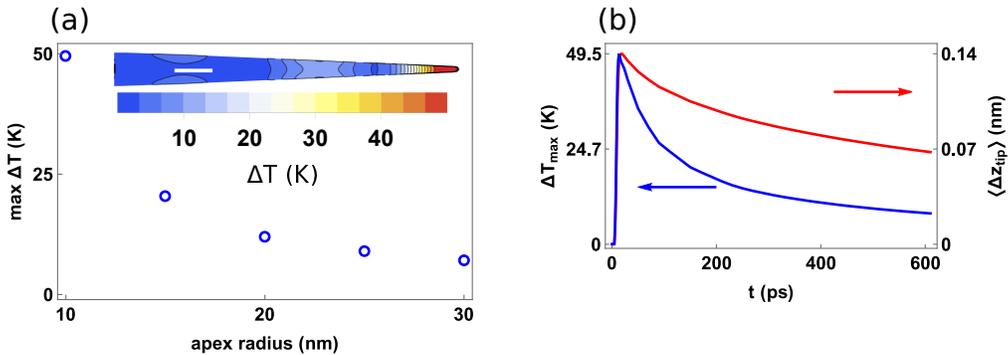}\]
\caption{\label{tipheat} \bf Heating the tip. \rm (a) Maximum temperature change of the plasmonic antennas as a function of apex radius. Inset: thermal profile for \SI{10}{nm} apex radius. Scale bar is \SI{100}{nm}. (b) The maximum temperature change of a \SI{10}{nm} tip apex (blue curve) and the corresponding apex thermal expansion (red curve) versus time, for an excitation pulse peaked at \SI{9}{ps}.}
\end{figure}

The inclusion of loss along the antenna heats up the tip significantly near the apex. We calculate tip heating for various apex radii placed above a metallic sample ($\sigma=$\SI{1e7}{S/m}), accounting for Ohmic and dielectric losses ($\sigma_{Au}$=\SI{4.5e7}{S/m}, $\epsilon_{r,Au}=-$22.855+1.4i). Gap distances between 1.3 and \SI{1.9}{nm} were chosen in the same manner as in Figs. 3(c) and 6(d), and SPPs are excited \SI{10}{\micro\metre} up the antenna by a \SI{1}{W} peak, \SI{3}{ps} duration laser pulse. The temperature increase in the antenna is largest for the smallest tip radii, where electric fields and charges are most strongly focused. The maximum temperature change, shown in Fig. \ref{tipheat}(a) for selected apex radii, is as large as \SI{50}{K} for a \SI{10}{nm} apex. As discussed in Section III\,B, this is low enough that there is not significant direct heat transfer between the apex and the sample compared to Ohmic heating excited in the sample. The inset in Fig. \ref{tipheat}(a) is the temperature change profile ($\Delta T$) for a \SI{10}{nm} apex \SI{4.5}{ps} after peak power input, when the maximum value of $\Delta T$ occurs. We plot the temperature along the lower \SI{1}{\micro\metre} of the tip to illustrate that heating occurs in bands corresponding to the coherent SPP wavelength along the tip, although heating is largest at the apex. These peaks get closer together as the SPPs are focused (discussed in Section S.III).

Heating the gold antenna causes thermal expansion; in Fig. \ref{tipheat}(b) we look at the magnitude of thermal expansion and its decay in relation to the maximum temperature in the tip. Expansion is considered in the vertical direction (i.e., towards the sample), accounting for significant heating in the lower \SI{1}{\micro\metre} of the antenna, and taking a thermal expansion coefficient of \SI{14e-6}{/K}\cite{Nix1941}. We find that expansion occurs mainly in the apex, and plot in Fig. \ref{tipheat}(b) the spatial average of vertical expansion throughout the apex region $\avg{\Delta z_{tip}}$, here for a \SI{10}{nm} apex radius up to \SI{610}{ps} (\SI{601}{ps} after peak input power). On the same plot, we consider the maximum temperature in the tip over the same range (in blue). Interestingly, the peak thermal expansion occurs at \SI{18}{ps}, \SI{4.5}{ps} after the maximum temperature (calculations were performed in \SI{1.5}{ps} steps), due to thermal diffusion throughout the tip. Furthermore, the thermal expansion decays much slower than the temperature change. Fitting each curve to two exponential decays, the long-term behavior of $\avg{\Delta z_{tip}}$ follows a characteristic decay time of \SI{1225+-23}{ps}, whereas the decay time of $\Delta T_{max}$ is \SI{602+-21}{ps}. Experimentally, laser pulses are spaced \SI{13}{ns} apart, thus there will be no residual thermal expansion or temperature rise between pulses. While the thermal expansion is non-zero, it is well below both the gap distance and the apex radius. We do not expect the expansion to interfere with operation of a scan probe microscope, where the apex will be at least \SI{1.3}{nm} from the sample. Because peak expansion does not occur until after the \SI{3}{ps} pulse has decayed, it will not significantly alter the tip-sample coupling.

\section{\label{sppprop}SPP Propagation Length}
In our numerical calculations we treat the gold antenna as lossless ($\sigma=0$, $Im[\epsilon_r]=0$) to investigate the sample thermal gradient without convolving it with effects of the SPP propagation. Here we determine the effective propagation length of SPPs along a tapered gold antenna, accounting for both Ohmic and dielectric losses. Field solutions along a tapered antenna are quantitatively different from those on a flat surface\cite{Babadjanyan2000}, and consequently the optical properties are modified. SPPs may exist at the boundary between vacuum and a metal of complex dielectric constant $\tilde{\epsilon}_m$, with effective refractive index of\cite{Raether1988}
\begin{equation}
\label{nflat}
\tilde n_{flat}=\sqrt{\frac{\tilde \epsilon_m}{\tilde\epsilon_m+1}}.
\end{equation}
Nanofocusing effects become important where $k_0R$, the product of the free-space wavenumber and the local antenna radius, is $<1$, i.e., where we can no longer treat the antenna surface as flat. In this regime,
\begin{equation}
\label{nrad}
\tilde n_{radius}(R)=\frac{1}{k_0R}\sqrt{-2/\tilde\epsilon_m}\bb{\log\sqrt{-4\tilde\epsilon_m}-\gamma}^{-1},
\end{equation}
where $\gamma\approx0.57721$ is the Euler constant\cite{Stockman2004}. The imaginary parts of Eq. \ref{nflat} and \ref{nrad} give us the dielectric absorption loss $\kappa_\epsilon(R)$, which is a function of local radius. Usually we ignore dielectric loss in metals and only consider Ohmic loss\cite{Staelin1998}, but the divergence of $\tilde n$ at small antenna radii suggests that dielectric loss is important in plasmonic nanofocusing.

\begin{figure}
\[\includegraphics[width=13.3cm]{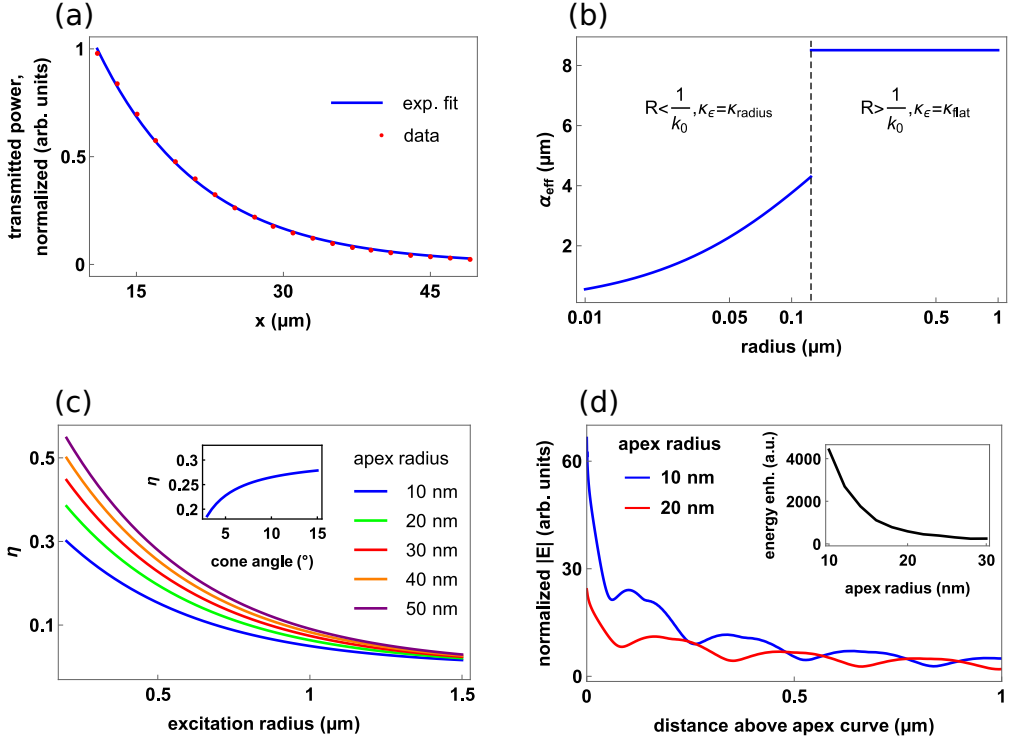}\]
\caption{\label{fig-proplength}\bf Propagation length and nanofocusing. \rm (a) Data and fit of power transmitted by SPPs along increasing lengths of a Au-air interface. (b) Effective propagation length, calculated from (a) and Eq. \ref{keff}-\ref{alphaeff}. (c) Remaining SPP power $\eta$, calculated with Eq. \ref{proplosseq}, for selected apex radii and SPP excitation distances from the apex. Inset: $\eta$ for varying cone angle, excited \SI{10}{\micro\metre} from the apex. (d) Electric field along the antenna edge for \SI{10}{nm} and \SI{20}{nm} apex radii. Inset: energy density enhancement versus tip radius.}
\end{figure}

We perform numerical calculations to determine conductive loss by exciting SPPs along a flat gold-air boundary ($\sigma_{Au}=$\SI{4.5e7}{S/m}, $\epsilon_{r,Au}=-22.855$) and measuring the power at the end of the surface for various propagation distances. By fitting these results to an exponential decay, shown in Fig. \ref{fig-proplength}(a), we extract a conductive loss constant of $\kappa_\sigma=0.012$, which is radius-independent.

The effective absorption loss is thus
\begin{equation}
\label{keff}
\kappa_{eff}=\kappa_\epsilon+\kappa_\sigma.
\end{equation}
In order to compute propagation loss for various antenna lengths, we make the assumption that
\begin{equation}
\kappa_\epsilon=\left\{\begin{array}{ll}\kappa_{flat},&k_0R>1\\\kappa_{radius},&k_0R<1\\\end{array}\right.,
\end{equation}
where $1/k_0=$\SI{124}{nm} for \SI{780}{nm} light. As shown in Fig. \ref{fig-proplength}(b), when the antenna radius is $<1/k_0$, the divergent behavior of $\tilde n_{radius}$ drives the effective propagation length,
\begin{equation}
\label{alphaeff}
\alpha_{eff}=\frac{1}{k_0\kappa_{eff}},
\end{equation}
to smaller and smaller values.

In Fig. \ref{fig-proplength}(c) we plot the energy remaining in the SPP wave when it reaches the apex, for apex radii of \SIrange{10}{50}{nm}, against excitation radius (antenna radius where light is incident) by evaluating\cite{Stockman2004}
\begin{equation}
\label{proplosseq}
\eta=\exp\bb{{\int_{z_{excite}}^{z_{apex}}\kappa_{eff}\pp{z\sin3^\circ}\,dz}},
\end{equation}
where $z$ is the length along the antenna surface and $z\sin3^\circ$ is the local radius for a cone half-angle of \ang{3}. Exciting closer to the apex leads to less loss, but experimentally we desire to decouple the multiple-micron wide laser spot from the apex to avoid direct excitation. This requires an excitation height of more than about \SI{5}{\micro\metre} (\SI{262}{nm} radius). Further constraints for placing diffraction gratings on antennas mean a more realistic height of \SI{10}{\micro\metre} (\SI{523}{nm} radius). For the apex radii we use in our heating calculations, an excitation height of \SI{10}{\micro\metre} leads to propagation loss of around 80\%, thus we scale down our input power by a factor of five for loss and heating calculations. Inset in Fig. \ref{fig-proplength}(c), we calculate loss as a function of cone angle by evaluating Eq. \ref{proplosseq} for \ang{1} to \ang{15}, starting \SI{10}{\micro\metre} from the apex. Experimentally, cone angles may be larger than \ang{3}, but this does not have a large effect on SPP loss.

Although Ohmic losses are significant along the antenna, the tapered geometry also increases the electric field\cite{Babadjanyan2000}, leading to a local enhancement of the energy density. In Fig. \ref{fig-proplength}(d) we look at electric field along the antenna edge, and energy density enhancement at the apex (inset), with the same antenna geometry discussed in Section II. We consider the antenna without a sample to demonstrate SPP focusing effects in isolation. The blue and red lines are electric field magnitude, $\abs{\v E}$, plotted up to \SI{1}{\micro\metre} away from where the apex begins (i.e., where the antenna stops being conical), for \SI{10}{nm} and \SI{20}{nm} apex radii, respectively. The field magnitude is normalized by the magnitude at the point where the SPPs are launched by the diffraction grating (here \SI{10}{\micro\metre} away from the apex). Inset is energy density enhancement for apex radii of \SIrange{10}{30}{nm}, showing larger enhancement for smaller radii. 

\begin{figure}[h]
\[\includegraphics[width=13.3cm]{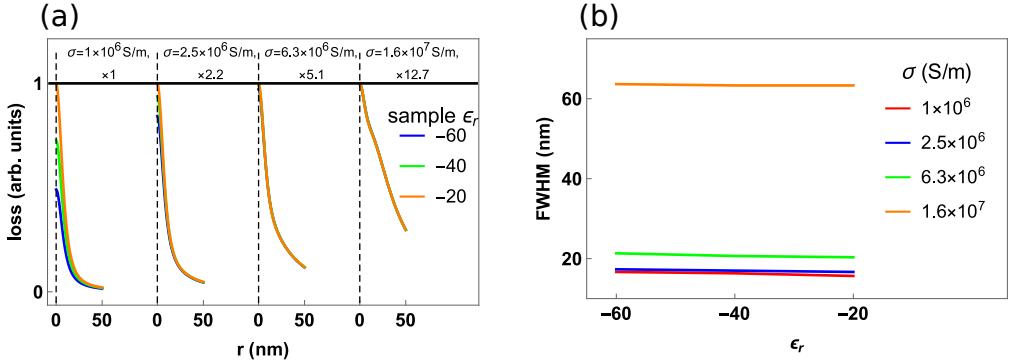}\]
\caption{\label{fig-sampleer} \bf Sample dielectric constant. \rm (a) Loss profiles for varying sample $\epsilon_r$ at different sample conductivities. Each panel is normalized by its respective peak value for ease of visualization. (b) The FWHM of profiles in (a). Variations in (b) are smaller than the mesh size used for calculations, and are negligible.}
\end{figure}

\section{\label{sampleer}Effect of dielectric constant on losses in metal sample}

The dielectric constant of the metal sample has a trivial effect on loss for realistic sample conductivity. In Fig. \ref{fig-sampleer}(a), we plot loss profiles in samples of selected conductivities, while varying the permittivity. For $\sigma=$\SI{1e6}{S/m}, varying the relative permittivity $\epsilon_r$ over a realistic range for metals at near-IR frequencies changes the magnitude of electromagnetic loss. As the conductivity approaches more realistic metal values, of $\sim$ \SI{5e6}{S/m}, there is negligible dependence on permittivity, as demonstrated by the overlap of loss profiles in Fig. \ref{fig-sampleer}(a). While the magnitude is affected by permittivity for low conductivity, the width of the loss profile is not. The FWHM for the profiles in Fig. \ref{fig-sampleer}(a) are shown in Fig. \ref{fig-sampleer}(b).

\pagebreak
\bibliography{arvix-paper}

\begin{thebibliography}{47}%
\makeatletter
\providecommand \@ifxundefined [1]{%
 \@ifx{#1\undefined}
}%
\providecommand \@ifnum [1]{%
 \ifnum #1\expandafter \@firstoftwo
 \else \expandafter \@secondoftwo
 \fi
}%
\providecommand \@ifx [1]{%
 \ifx #1\expandafter \@firstoftwo
 \else \expandafter \@secondoftwo
 \fi
}%
\providecommand \natexlab [1]{#1}%
\providecommand \enquote  [1]{``#1''}%
\providecommand \bibnamefont  [1]{#1}%
\providecommand \bibfnamefont [1]{#1}%
\providecommand \citenamefont [1]{#1}%
\providecommand \href@noop [0]{\@secondoftwo}%
\providecommand \href [0]{\begingroup \@sanitize@url \@href}%
\providecommand \@href[1]{\@@startlink{#1}\@@href}%
\providecommand \@@href[1]{\endgroup#1\@@endlink}%
\providecommand \@sanitize@url [0]{\catcode `\\12\catcode `\$12\catcode
  `\&12\catcode `\#12\catcode `\^12\catcode `\_12\catcode `\%12\relax}%
\providecommand \@@startlink[1]{}%
\providecommand \@@endlink[0]{}%
\providecommand \url  [0]{\begingroup\@sanitize@url \@url }%
\providecommand \@url [1]{\endgroup\@href {#1}{\urlprefix }}%
\providecommand \urlprefix  [0]{URL }%
\providecommand \Eprint [0]{\href }%
\providecommand \doibase [0]{http://dx.doi.org/}%
\providecommand \selectlanguage [0]{\@gobble}%
\providecommand \bibinfo  [0]{\@secondoftwo}%
\providecommand \bibfield  [0]{\@secondoftwo}%
\providecommand \translation [1]{[#1]}%
\providecommand \BibitemOpen [0]{}%
\providecommand \bibitemStop [0]{}%
\providecommand \bibitemNoStop [0]{.\EOS\space}%
\providecommand \EOS [0]{\spacefactor3000\relax}%
\providecommand \BibitemShut  [1]{\csname bibitem#1\endcsname}%
\let\auto@bib@innerbib\@empty
\bibitem [{\citenamefont {Wolf}\ \emph {et~al.}(2001)\citenamefont {Wolf},
  \citenamefont {Awschalom}, \citenamefont {Buhrman}, \citenamefont {Daughton},
  \citenamefont {von Moln{\'{a}}r}, \citenamefont {Roukes}, \citenamefont
  {Chtchelkanova},\ and\ \citenamefont {Treger}}]{Wolf2001}%
  \BibitemOpen
  \bibfield  {author} {\bibinfo {author} {\bibfnamefont {S.~A.}\ \bibnamefont
  {Wolf}}, \bibinfo {author} {\bibfnamefont {D.~D.}\ \bibnamefont {Awschalom}},
  \bibinfo {author} {\bibfnamefont {R.~A.}\ \bibnamefont {Buhrman}}, \bibinfo
  {author} {\bibfnamefont {J.~M.}\ \bibnamefont {Daughton}}, \bibinfo {author}
  {\bibfnamefont {S.}~\bibnamefont {von Moln{\'{a}}r}}, \bibinfo {author}
  {\bibfnamefont {M.~L.}\ \bibnamefont {Roukes}}, \bibinfo {author}
  {\bibfnamefont {A.~Y.}\ \bibnamefont {Chtchelkanova}}, \ and\ \bibinfo
  {author} {\bibfnamefont {D.~M.}\ \bibnamefont {Treger}},\ }\href
  {http://science.sciencemag.org/content/294/5546/1488.full} {\bibfield
  {journal} {\bibinfo  {journal} {Science}\ }\textbf {\bibinfo {volume} {294}}
  (\bibinfo {year} {2001})}\BibitemShut {NoStop}%
\bibitem [{\citenamefont {Challener}\ \emph {et~al.}(2009)\citenamefont
  {Challener}, \citenamefont {Peng}, \citenamefont {Karns}, \citenamefont
  {Peng}, \citenamefont {Peng}, \citenamefont {Yang}, \citenamefont {Zhu},
  \citenamefont {Gokemeijer}, \citenamefont {Hsia}, \citenamefont {Ju},
  \citenamefont {Rottmayer}, \citenamefont {Seigler},\ and\ \citenamefont
  {Gage}}]{Challener2009}%
  \BibitemOpen
  \bibfield  {author} {\bibinfo {author} {\bibnamefont {Challener}}, \bibinfo
  {author} {\bibfnamefont {C.}~\bibnamefont {Peng}}, \bibinfo {author}
  {\bibfnamefont {I.}~\bibnamefont {Karns}}, \bibinfo {author} {\bibfnamefont
  {W.}~\bibnamefont {Peng}}, \bibinfo {author} {\bibfnamefont {Y.}~\bibnamefont
  {Peng}}, \bibinfo {author} {\bibfnamefont {X.}~\bibnamefont {Yang}}, \bibinfo
  {author} {\bibfnamefont {X.}~\bibnamefont {Zhu}}, \bibinfo {author}
  {\bibfnamefont {N.~J.}\ \bibnamefont {Gokemeijer}}, \bibinfo {author}
  {\bibfnamefont {Y.-T.}\ \bibnamefont {Hsia}}, \bibinfo {author}
  {\bibfnamefont {G.}~\bibnamefont {Ju}}, \bibinfo {author} {\bibfnamefont
  {R.~E.}\ \bibnamefont {Rottmayer}}, \bibinfo {author} {\bibfnamefont {M.~A.}\
  \bibnamefont {Seigler}}, \ and\ \bibinfo {author} {\bibfnamefont {E.~C.}\
  \bibnamefont {Gage}},\ }\href {\doibase 10.1038/NPHOTON.2009.26} {\bibfield
  {journal} {\bibinfo  {journal} {Nat. Photonics}\ }\textbf {\bibinfo {volume}
  {3}},\ \bibinfo {pages} {220} (\bibinfo {year} {2009})}\BibitemShut {NoStop}%
\bibitem [{\citenamefont {Beach}\ \emph {et~al.}(2005)\citenamefont {Beach},
  \citenamefont {Nistor}, \citenamefont {Knutson}, \citenamefont {Tsoi},\ and\
  \citenamefont {Erskine}}]{beach2005}%
  \BibitemOpen
  \bibfield  {author} {\bibinfo {author} {\bibfnamefont {G.~S.~D.}\
  \bibnamefont {Beach}}, \bibinfo {author} {\bibfnamefont {C.}~\bibnamefont
  {Nistor}}, \bibinfo {author} {\bibfnamefont {C.}~\bibnamefont {Knutson}},
  \bibinfo {author} {\bibfnamefont {M.}~\bibnamefont {Tsoi}}, \ and\ \bibinfo
  {author} {\bibfnamefont {J.~L.}\ \bibnamefont {Erskine}},\ }\href {\doibase
  10.1038/nmat1477} {\bibfield  {journal} {\bibinfo  {journal} {Nat. Mater.}\
  }\textbf {\bibinfo {volume} {4}},\ \bibinfo {pages} {741} (\bibinfo {year}
  {2005})}\BibitemShut {NoStop}%
\bibitem [{\citenamefont {Yamaguchi}\ \emph {et~al.}(2004)\citenamefont
  {Yamaguchi}, \citenamefont {Ono}, \citenamefont {Nasu}, \citenamefont
  {Miyake}, \citenamefont {Mibu},\ and\ \citenamefont
  {Shinjo}}]{yamaguchi2004}%
  \BibitemOpen
  \bibfield  {author} {\bibinfo {author} {\bibfnamefont {A.}~\bibnamefont
  {Yamaguchi}}, \bibinfo {author} {\bibfnamefont {T.}~\bibnamefont {Ono}},
  \bibinfo {author} {\bibfnamefont {S.}~\bibnamefont {Nasu}}, \bibinfo {author}
  {\bibfnamefont {K.}~\bibnamefont {Miyake}}, \bibinfo {author} {\bibfnamefont
  {K.}~\bibnamefont {Mibu}}, \ and\ \bibinfo {author} {\bibfnamefont
  {T.}~\bibnamefont {Shinjo}},\ }\href {\doibase 10.1103/PhysRevLett.92.077205}
  {\bibfield  {journal} {\bibinfo  {journal} {Phys. Rev. Lett.}\ }\textbf
  {\bibinfo {volume} {92}},\ \bibinfo {pages} {077205} (\bibinfo {year}
  {2004})}\BibitemShut {NoStop}%
\bibitem [{\citenamefont {Mangin}\ \emph {et~al.}(2006)\citenamefont {Mangin},
  \citenamefont {Ravelosona}, \citenamefont {Katine}, \citenamefont {Carey},
  \citenamefont {Terris},\ and\ \citenamefont {Fullerton}}]{mangin2006}%
  \BibitemOpen
  \bibfield  {author} {\bibinfo {author} {\bibfnamefont {S.}~\bibnamefont
  {Mangin}}, \bibinfo {author} {\bibfnamefont {D.}~\bibnamefont {Ravelosona}},
  \bibinfo {author} {\bibfnamefont {J.~A.}\ \bibnamefont {Katine}}, \bibinfo
  {author} {\bibfnamefont {M.~J.}\ \bibnamefont {Carey}}, \bibinfo {author}
  {\bibfnamefont {B.~D.}\ \bibnamefont {Terris}}, \ and\ \bibinfo {author}
  {\bibfnamefont {E.~E.}\ \bibnamefont {Fullerton}},\ }\href {\doibase
  10.1038/nmat1595} {\bibfield  {journal} {\bibinfo  {journal} {Nat. Mater.}\
  }\textbf {\bibinfo {volume} {5}},\ \bibinfo {pages} {210} (\bibinfo {year}
  {2006})}\BibitemShut {NoStop}%
\bibitem [{\citenamefont {Madami}\ \emph {et~al.}(2011)\citenamefont {Madami},
  \citenamefont {Bonetti}, \citenamefont {Consolo}, \citenamefont {Tacchi},
  \citenamefont {Carlotti}, \citenamefont {Gubbiotti}, \citenamefont {Mancoff},
  \citenamefont {Yar},\ and\ \citenamefont {{\r{A}}kerman}}]{madami2011}%
  \BibitemOpen
  \bibfield  {author} {\bibinfo {author} {\bibfnamefont {M.}~\bibnamefont
  {Madami}}, \bibinfo {author} {\bibfnamefont {S.}~\bibnamefont {Bonetti}},
  \bibinfo {author} {\bibfnamefont {G.}~\bibnamefont {Consolo}}, \bibinfo
  {author} {\bibfnamefont {S.}~\bibnamefont {Tacchi}}, \bibinfo {author}
  {\bibfnamefont {G.}~\bibnamefont {Carlotti}}, \bibinfo {author}
  {\bibfnamefont {G.}~\bibnamefont {Gubbiotti}}, \bibinfo {author}
  {\bibfnamefont {F.~B.}\ \bibnamefont {Mancoff}}, \bibinfo {author}
  {\bibfnamefont {M.~A.}\ \bibnamefont {Yar}}, \ and\ \bibinfo {author}
  {\bibfnamefont {J.}~\bibnamefont {{\r{A}}kerman}},\ }\href {\doibase
  10.1038/nnano.2011.140} {\bibfield  {journal} {\bibinfo  {journal} {Nat.
  Nanotechnol.}\ }\textbf {\bibinfo {volume} {6}},\ \bibinfo {pages} {635}
  (\bibinfo {year} {2011})}\BibitemShut {NoStop}%
\bibitem [{\citenamefont {Katine}\ \emph {et~al.}(2000)\citenamefont {Katine},
  \citenamefont {Albert}, \citenamefont {Buhrman}, \citenamefont {Myers},\ and\
  \citenamefont {Ralph}}]{katine2000}%
  \BibitemOpen
  \bibfield  {author} {\bibinfo {author} {\bibfnamefont {J.~A.}\ \bibnamefont
  {Katine}}, \bibinfo {author} {\bibfnamefont {F.}~\bibnamefont {Albert}},
  \bibinfo {author} {\bibfnamefont {R.}~\bibnamefont {Buhrman}}, \bibinfo
  {author} {\bibfnamefont {E.}~\bibnamefont {Myers}}, \ and\ \bibinfo {author}
  {\bibfnamefont {D.}~\bibnamefont {Ralph}},\ }\href {\doibase
  10.1103/PhysRevLett.84.3149} {\bibfield  {journal} {\bibinfo  {journal}
  {Phys. Rev. Lett.}\ }\textbf {\bibinfo {volume} {84}},\ \bibinfo {pages}
  {3149} (\bibinfo {year} {2000})}\BibitemShut {NoStop}%
\bibitem [{\citenamefont {Bartell}\ \emph {et~al.}(2015)\citenamefont
  {Bartell}, \citenamefont {Ngai}, \citenamefont {Leng},\ and\ \citenamefont
  {Fuchs}}]{bartell2015}%
  \BibitemOpen
  \bibfield  {author} {\bibinfo {author} {\bibfnamefont {J.~M.}\ \bibnamefont
  {Bartell}}, \bibinfo {author} {\bibfnamefont {D.~H.}\ \bibnamefont {Ngai}},
  \bibinfo {author} {\bibfnamefont {Z.}~\bibnamefont {Leng}}, \ and\ \bibinfo
  {author} {\bibfnamefont {G.~D.}\ \bibnamefont {Fuchs}},\ }\href {\doibase
  10.1038/ncomms9460} {\bibfield  {journal} {\bibinfo  {journal} {Nat.
  Commun.}\ }\textbf {\bibinfo {volume} {6}},\ \bibinfo {pages} {8460}
  (\bibinfo {year} {2015})}\BibitemShut {NoStop}%
\bibitem [{\citenamefont {Freeman}\ and\ \citenamefont
  {Choi}(2001)}]{freeman2001}%
  \BibitemOpen
  \bibfield  {author} {\bibinfo {author} {\bibfnamefont {M.~R.}\ \bibnamefont
  {Freeman}}\ and\ \bibinfo {author} {\bibfnamefont {B.~C.}\ \bibnamefont
  {Choi}},\ }\href {\doibase 10.1126/science.1065300} {\bibfield  {journal}
  {\bibinfo  {journal} {Science}\ }\textbf {\bibinfo {volume} {294}},\ \bibinfo
  {pages} {1484} (\bibinfo {year} {2001})}\BibitemShut {NoStop}%
\bibitem [{\citenamefont {Hubert}\ and\ \citenamefont
  {Schafer}(1998)}]{hubert1998}%
  \BibitemOpen
  \bibfield  {author} {\bibinfo {author} {\bibfnamefont {A.}~\bibnamefont
  {Hubert}}\ and\ \bibinfo {author} {\bibfnamefont {R.}~\bibnamefont
  {Schafer}},\ }\href {\doibase 10.1007/978-3-540-85054-0} {\emph {\bibinfo
  {title} {{Magnetic Domains}}}}\ (\bibinfo  {publisher} {Springer},\ \bibinfo
  {address} {Berlin, Heidelberg},\ \bibinfo {year} {1998})\BibitemShut
  {NoStop}%
\bibitem [{\citenamefont {Trunk}\ \emph {et~al.}(2001)\citenamefont {Trunk},
  \citenamefont {Redjdal}, \citenamefont {K{\'{a}}kay}, \citenamefont {Ruane},\
  and\ \citenamefont {Humphrey}}]{trunk2001}%
  \BibitemOpen
  \bibfield  {author} {\bibinfo {author} {\bibfnamefont {T.}~\bibnamefont
  {Trunk}}, \bibinfo {author} {\bibfnamefont {M.}~\bibnamefont {Redjdal}},
  \bibinfo {author} {\bibfnamefont {A.}~\bibnamefont {K{\'{a}}kay}}, \bibinfo
  {author} {\bibfnamefont {M.~F.}\ \bibnamefont {Ruane}}, \ and\ \bibinfo
  {author} {\bibfnamefont {F.~B.}\ \bibnamefont {Humphrey}},\ }\href {\doibase
  10.1063/1.1355357} {\bibfield  {journal} {\bibinfo  {journal} {J. Appl.
  Phys.}\ }\textbf {\bibinfo {volume} {89}},\ \bibinfo {pages} {7606} (\bibinfo
  {year} {2001})}\BibitemShut {NoStop}%
\bibitem [{\citenamefont {Allenspach}, \citenamefont {Stampanoni},\ and\
  \citenamefont {Bischof}(1990)}]{allenspach1990}%
  \BibitemOpen
  \bibfield  {author} {\bibinfo {author} {\bibfnamefont {R.}~\bibnamefont
  {Allenspach}}, \bibinfo {author} {\bibfnamefont {M.}~\bibnamefont
  {Stampanoni}}, \ and\ \bibinfo {author} {\bibfnamefont {A.}~\bibnamefont
  {Bischof}},\ }\href {\doibase 10.1103/PhysRevLett.65.3344} {\bibfield
  {journal} {\bibinfo  {journal} {Phys. Rev. Lett.}\ }\textbf {\bibinfo
  {volume} {65}},\ \bibinfo {pages} {3344} (\bibinfo {year}
  {1990})}\BibitemShut {NoStop}%
\bibitem [{\citenamefont {O'Brien}\ \emph {et~al.}(2009)\citenamefont
  {O'Brien}, \citenamefont {Petit}, \citenamefont {Zeng}, \citenamefont
  {Lewis}, \citenamefont {Sampaio}, \citenamefont {Jausovec}, \citenamefont
  {Read},\ and\ \citenamefont {Cowburn}}]{brien2009}%
  \BibitemOpen
  \bibfield  {author} {\bibinfo {author} {\bibfnamefont {L.}~\bibnamefont
  {O'Brien}}, \bibinfo {author} {\bibfnamefont {D.}~\bibnamefont {Petit}},
  \bibinfo {author} {\bibfnamefont {H.~T.}\ \bibnamefont {Zeng}}, \bibinfo
  {author} {\bibfnamefont {E.~R.}\ \bibnamefont {Lewis}}, \bibinfo {author}
  {\bibfnamefont {J.}~\bibnamefont {Sampaio}}, \bibinfo {author} {\bibfnamefont
  {A.~V.}\ \bibnamefont {Jausovec}}, \bibinfo {author} {\bibfnamefont {D.~E.}\
  \bibnamefont {Read}}, \ and\ \bibinfo {author} {\bibfnamefont {R.~P.}\
  \bibnamefont {Cowburn}},\ }\href {\doibase 10.1103/PhysRevLett.103.077206}
  {\bibfield  {journal} {\bibinfo  {journal} {Phys. Rev. Lett.}\ }\textbf
  {\bibinfo {volume} {103}},\ \bibinfo {pages} {077206} (\bibinfo {year}
  {2009})}\BibitemShut {NoStop}%
\bibitem [{\citenamefont {Acremann}\ \emph {et~al.}(2006)\citenamefont
  {Acremann}, \citenamefont {Strachan}, \citenamefont {Chembrolu},
  \citenamefont {Andrews}, \citenamefont {Tyliszczak}, \citenamefont {Katine},
  \citenamefont {Carey}, \citenamefont {Clemens}, \citenamefont {Siegmann},\
  and\ \citenamefont {St{\"{o}}hr}}]{acremann2006}%
  \BibitemOpen
  \bibfield  {author} {\bibinfo {author} {\bibfnamefont {Y.}~\bibnamefont
  {Acremann}}, \bibinfo {author} {\bibfnamefont {J.}~\bibnamefont {Strachan}},
  \bibinfo {author} {\bibfnamefont {V.}~\bibnamefont {Chembrolu}}, \bibinfo
  {author} {\bibfnamefont {S.}~\bibnamefont {Andrews}}, \bibinfo {author}
  {\bibfnamefont {T.}~\bibnamefont {Tyliszczak}}, \bibinfo {author}
  {\bibfnamefont {J.}~\bibnamefont {Katine}}, \bibinfo {author} {\bibfnamefont
  {M.}~\bibnamefont {Carey}}, \bibinfo {author} {\bibfnamefont
  {B.}~\bibnamefont {Clemens}}, \bibinfo {author} {\bibfnamefont
  {H.}~\bibnamefont {Siegmann}}, \ and\ \bibinfo {author} {\bibfnamefont
  {J.}~\bibnamefont {St{\"{o}}hr}},\ }\href {\doibase
  10.1103/PhysRevLett.96.217202} {\bibfield  {journal} {\bibinfo  {journal}
  {Phys. Rev. Lett.}\ }\textbf {\bibinfo {volume} {96}},\ \bibinfo {pages}
  {217202} (\bibinfo {year} {2006})}\BibitemShut {NoStop}%
\bibitem [{\citenamefont {St{\"{o}}hr}(1995)}]{stohr1995}%
  \BibitemOpen
  \bibfield  {author} {\bibinfo {author} {\bibfnamefont {J.}~\bibnamefont
  {St{\"{o}}hr}},\ }\href {\doibase 10.1016/0368-2048(95)02537-5} {\bibfield
  {journal} {\bibinfo  {journal} {J. Electron Spectros. Relat. Phenomena}\
  }\textbf {\bibinfo {volume} {75}},\ \bibinfo {pages} {253} (\bibinfo {year}
  {1995})}\BibitemShut {NoStop}%
\bibitem [{\citenamefont {Oppeneer}\ \emph {et~al.}(1992)\citenamefont
  {Oppeneer}, \citenamefont {Maurer}, \citenamefont {Sticht},\ and\
  \citenamefont {K{\"{u}}bler}}]{oppeneer1992}%
  \BibitemOpen
  \bibfield  {author} {\bibinfo {author} {\bibfnamefont {P.~M.}\ \bibnamefont
  {Oppeneer}}, \bibinfo {author} {\bibfnamefont {T.}~\bibnamefont {Maurer}},
  \bibinfo {author} {\bibfnamefont {J.}~\bibnamefont {Sticht}}, \ and\ \bibinfo
  {author} {\bibfnamefont {J.}~\bibnamefont {K{\"{u}}bler}},\ }\href {\doibase
  10.1103/PhysRevB.45.10924} {\bibfield  {journal} {\bibinfo  {journal} {Phys.
  Rev. B}\ }\textbf {\bibinfo {volume} {45}},\ \bibinfo {pages} {10924}
  (\bibinfo {year} {1992})}\BibitemShut {NoStop}%
\bibitem [{\citenamefont {Hecht}(2002)}]{hecht2002}%
  \BibitemOpen
  \bibfield  {author} {\bibinfo {author} {\bibfnamefont {E.}~\bibnamefont
  {Hecht}},\ }\href@noop {} {\emph {\bibinfo {title} {{Optics}}}}\ (\bibinfo
  {publisher} {Addison-Wesley},\ \bibinfo {year} {2002})\BibitemShut {NoStop}%
\bibitem [{\citenamefont {Guo}\ \emph {et~al.}(2015)\citenamefont {Guo},
  \citenamefont {Bartell}, \citenamefont {Ngai},\ and\ \citenamefont
  {Fuchs}}]{Guo2015}%
  \BibitemOpen
  \bibfield  {author} {\bibinfo {author} {\bibfnamefont {F.}~\bibnamefont
  {Guo}}, \bibinfo {author} {\bibfnamefont {J.}~\bibnamefont {Bartell}},
  \bibinfo {author} {\bibfnamefont {D.}~\bibnamefont {Ngai}}, \ and\ \bibinfo
  {author} {\bibfnamefont {G.}~\bibnamefont {Fuchs}},\ }\href {\doibase
  10.1103/PhysRevApplied.4.044004} {\bibfield  {journal} {\bibinfo  {journal}
  {Phys. Rev. Appl.}\ }\textbf {\bibinfo {volume} {4}},\ \bibinfo {pages}
  {044004} (\bibinfo {year} {2015})}\BibitemShut {NoStop}%
\bibitem [{\citenamefont {Bartell}\ \emph {et~al.}(2017)\citenamefont
  {Bartell}, \citenamefont {Jermain}, \citenamefont {Aradhya}, \citenamefont
  {Brangham}, \citenamefont {Yang}, \citenamefont {Ralph},\ and\ \citenamefont
  {Fuchs}}]{Bartell2017}%
  \BibitemOpen
  \bibfield  {author} {\bibinfo {author} {\bibfnamefont {J.~M.}\ \bibnamefont
  {Bartell}}, \bibinfo {author} {\bibfnamefont {C.~L.}\ \bibnamefont
  {Jermain}}, \bibinfo {author} {\bibfnamefont {S.~V.}\ \bibnamefont
  {Aradhya}}, \bibinfo {author} {\bibfnamefont {J.~T.}\ \bibnamefont
  {Brangham}}, \bibinfo {author} {\bibfnamefont {F.}~\bibnamefont {Yang}},
  \bibinfo {author} {\bibfnamefont {D.~C.}\ \bibnamefont {Ralph}}, \ and\
  \bibinfo {author} {\bibfnamefont {G.~D.}\ \bibnamefont {Fuchs}},\ }\href
  {\doibase 10.1103/PhysRevApplied.7.044004} {\bibfield  {journal} {\bibinfo
  {journal} {Phys. Rev. Appl.}\ }\textbf {\bibinfo {volume} {7}},\ \bibinfo
  {pages} {044004} (\bibinfo {year} {2017})}\BibitemShut {NoStop}%
\bibitem [{\citenamefont {Weiler}\ \emph {et~al.}(2012)\citenamefont {Weiler},
  \citenamefont {Althammer}, \citenamefont {Czeschka}, \citenamefont {Huebl},
  \citenamefont {Wagner}, \citenamefont {Opel}, \citenamefont {Imort},
  \citenamefont {Reiss}, \citenamefont {Thomas}, \citenamefont {Gross},\ and\
  \citenamefont {Goennenwein}}]{Weiler2012}%
  \BibitemOpen
  \bibfield  {author} {\bibinfo {author} {\bibfnamefont {M.}~\bibnamefont
  {Weiler}}, \bibinfo {author} {\bibfnamefont {M.}~\bibnamefont {Althammer}},
  \bibinfo {author} {\bibfnamefont {F.~D.}\ \bibnamefont {Czeschka}}, \bibinfo
  {author} {\bibfnamefont {H.}~\bibnamefont {Huebl}}, \bibinfo {author}
  {\bibfnamefont {M.~S.}\ \bibnamefont {Wagner}}, \bibinfo {author}
  {\bibfnamefont {M.}~\bibnamefont {Opel}}, \bibinfo {author} {\bibfnamefont
  {I.-M.}\ \bibnamefont {Imort}}, \bibinfo {author} {\bibfnamefont
  {G.}~\bibnamefont {Reiss}}, \bibinfo {author} {\bibfnamefont
  {A.}~\bibnamefont {Thomas}}, \bibinfo {author} {\bibfnamefont
  {R.}~\bibnamefont {Gross}}, \ and\ \bibinfo {author} {\bibfnamefont
  {S.~T.~B.}\ \bibnamefont {Goennenwein}},\ }\href {\doibase
  10.1103/PhysRevLett.108.106602} {\bibfield  {journal} {\bibinfo  {journal}
  {Phys. Rev. Lett.}\ }\textbf {\bibinfo {volume} {108}},\ \bibinfo {pages}
  {106602} (\bibinfo {year} {2012})}\BibitemShut {NoStop}%
\bibitem [{\citenamefont {von Bieren}\ \emph {et~al.}(2013)\citenamefont {von
  Bieren}, \citenamefont {Brandl}, \citenamefont {Grundler},\ and\
  \citenamefont {Ansermet}}]{VonBieren2013}%
  \BibitemOpen
  \bibfield  {author} {\bibinfo {author} {\bibfnamefont {A.}~\bibnamefont {von
  Bieren}}, \bibinfo {author} {\bibfnamefont {F.}~\bibnamefont {Brandl}},
  \bibinfo {author} {\bibfnamefont {D.}~\bibnamefont {Grundler}}, \ and\
  \bibinfo {author} {\bibfnamefont {J.-P.}\ \bibnamefont {Ansermet}},\ }\href
  {\doibase 10.1063/1.4789974} {\bibfield  {journal} {\bibinfo  {journal}
  {Appl. Phys. Lett.}\ }\textbf {\bibinfo {volume} {102}},\ \bibinfo {pages}
  {052408} (\bibinfo {year} {2013})}\BibitemShut {NoStop}%
\bibitem [{\citenamefont {Schmidt}\ \emph {et~al.}(2012)\citenamefont
  {Schmidt}, \citenamefont {Piglosiewicz}, \citenamefont {Sadiq}, \citenamefont
  {Shirdel}, \citenamefont {Lee}, \citenamefont {Vasa}, \citenamefont {Park},
  \citenamefont {Kim},\ and\ \citenamefont {Lienau}}]{Schmidt2012}%
  \BibitemOpen
  \bibfield  {author} {\bibinfo {author} {\bibfnamefont {S.}~\bibnamefont
  {Schmidt}}, \bibinfo {author} {\bibfnamefont {B.}~\bibnamefont
  {Piglosiewicz}}, \bibinfo {author} {\bibfnamefont {D.}~\bibnamefont {Sadiq}},
  \bibinfo {author} {\bibfnamefont {J.}~\bibnamefont {Shirdel}}, \bibinfo
  {author} {\bibfnamefont {J.~S.}\ \bibnamefont {Lee}}, \bibinfo {author}
  {\bibfnamefont {P.}~\bibnamefont {Vasa}}, \bibinfo {author} {\bibfnamefont
  {N.}~\bibnamefont {Park}}, \bibinfo {author} {\bibfnamefont {D.~S.}\
  \bibnamefont {Kim}}, \ and\ \bibinfo {author} {\bibfnamefont
  {C.}~\bibnamefont {Lienau}},\ }\href {\doibase 10.1021/nn301121h} {\bibfield
  {journal} {\bibinfo  {journal} {ACS Nano}\ }\textbf {\bibinfo {volume} {6}},\
  \bibinfo {pages} {6040} (\bibinfo {year} {2012})}\BibitemShut {NoStop}%
\bibitem [{\citenamefont {Babadjanyan}, \citenamefont {Margaryan},\ and\
  \citenamefont {Nerkararyan}(2000)}]{Babadjanyan2000}%
  \BibitemOpen
  \bibfield  {author} {\bibinfo {author} {\bibfnamefont {A.~J.}\ \bibnamefont
  {Babadjanyan}}, \bibinfo {author} {\bibfnamefont {N.~L.}\ \bibnamefont
  {Margaryan}}, \ and\ \bibinfo {author} {\bibfnamefont {K.~V.}\ \bibnamefont
  {Nerkararyan}},\ }\href {\doibase doi:10.1063/1.372414} {\bibfield  {journal}
  {\bibinfo  {journal} {J. Appl. Phys.}\ }\textbf {\bibinfo {volume} {87}},\
  \bibinfo {pages} {3785} (\bibinfo {year} {2000})}\BibitemShut {NoStop}%
\bibitem [{\citenamefont {Ropers}\ \emph {et~al.}(2007)\citenamefont {Ropers},
  \citenamefont {Neacsu}, \citenamefont {Elsaesser}, \citenamefont {Albrecht},
  \citenamefont {Raschke},\ and\ \citenamefont {Lienau}}]{Ropers2007}%
  \BibitemOpen
  \bibfield  {author} {\bibinfo {author} {\bibfnamefont {C.}~\bibnamefont
  {Ropers}}, \bibinfo {author} {\bibfnamefont {C.~C.}\ \bibnamefont {Neacsu}},
  \bibinfo {author} {\bibfnamefont {T.}~\bibnamefont {Elsaesser}}, \bibinfo
  {author} {\bibfnamefont {M.}~\bibnamefont {Albrecht}}, \bibinfo {author}
  {\bibfnamefont {M.~B.}\ \bibnamefont {Raschke}}, \ and\ \bibinfo {author}
  {\bibfnamefont {C.}~\bibnamefont {Lienau}},\ }\href {\doibase
  10.1021/nl071340m} {\bibfield  {journal} {\bibinfo  {journal} {Nano Lett.}\
  }\textbf {\bibinfo {volume} {7}},\ \bibinfo {pages} {2784} (\bibinfo {year}
  {2007})}\BibitemShut {NoStop}%
\bibitem [{\citenamefont {Neacsu}\ \emph {et~al.}(2010)\citenamefont {Neacsu},
  \citenamefont {Berweger}, \citenamefont {Olmon}, \citenamefont {Saraf},
  \citenamefont {Ropers},\ and\ \citenamefont {Raschke}}]{neacsu2010}%
  \BibitemOpen
  \bibfield  {author} {\bibinfo {author} {\bibfnamefont {C.~C.}\ \bibnamefont
  {Neacsu}}, \bibinfo {author} {\bibfnamefont {S.}~\bibnamefont {Berweger}},
  \bibinfo {author} {\bibfnamefont {R.~L.}\ \bibnamefont {Olmon}}, \bibinfo
  {author} {\bibfnamefont {L.~V.}\ \bibnamefont {Saraf}}, \bibinfo {author}
  {\bibfnamefont {C.}~\bibnamefont {Ropers}}, \ and\ \bibinfo {author}
  {\bibfnamefont {M.~B.}\ \bibnamefont {Raschke}},\ }\href {\doibase
  10.1021/nl903574a} {\bibfield  {journal} {\bibinfo  {journal} {Nano Lett.}\
  }\textbf {\bibinfo {volume} {10}},\ \bibinfo {pages} {592} (\bibinfo {year}
  {2010})}\BibitemShut {NoStop}%
\bibitem [{\citenamefont {M{\"{u}}ller}\ \emph {et~al.}(2016)\citenamefont
  {M{\"{u}}ller}, \citenamefont {Kravtsov}, \citenamefont {Paarmann},
  \citenamefont {Raschke},\ and\ \citenamefont {Ernstorfer}}]{Muller2016}%
  \BibitemOpen
  \bibfield  {author} {\bibinfo {author} {\bibfnamefont {M.}~\bibnamefont
  {M{\"{u}}ller}}, \bibinfo {author} {\bibfnamefont {V.}~\bibnamefont
  {Kravtsov}}, \bibinfo {author} {\bibfnamefont {A.}~\bibnamefont {Paarmann}},
  \bibinfo {author} {\bibfnamefont {M.~B.}\ \bibnamefont {Raschke}}, \ and\
  \bibinfo {author} {\bibfnamefont {R.}~\bibnamefont {Ernstorfer}},\ }\href
  {\doibase 10.1021/acsphotonics.5b00710} {\bibfield  {journal} {\bibinfo
  {journal} {ACS Photonics}\ }\textbf {\bibinfo {volume} {3}},\ \bibinfo
  {pages} {611} (\bibinfo {year} {2016})}\BibitemShut {NoStop}%
\bibitem [{\citenamefont {Ren}, \citenamefont {Picardi},\ and\ \citenamefont
  {Pettinger}(2004)}]{ren2004}%
  \BibitemOpen
  \bibfield  {author} {\bibinfo {author} {\bibfnamefont {B.}~\bibnamefont
  {Ren}}, \bibinfo {author} {\bibfnamefont {G.}~\bibnamefont {Picardi}}, \ and\
  \bibinfo {author} {\bibfnamefont {B.}~\bibnamefont {Pettinger}},\ }\href
  {\doibase 10.1063/1.1688442} {\bibfield  {journal} {\bibinfo  {journal} {Rev.
  Sci. Instrum.}\ }\textbf {\bibinfo {volume} {75}},\ \bibinfo {pages} {837}
  (\bibinfo {year} {2004})}\BibitemShut {NoStop}%
\bibitem [{\citenamefont {Liao}\ and\ \citenamefont {Wokaun}(1982)}]{liao1982}%
  \BibitemOpen
  \bibfield  {author} {\bibinfo {author} {\bibfnamefont {P.~F.}\ \bibnamefont
  {Liao}}\ and\ \bibinfo {author} {\bibfnamefont {A.}~\bibnamefont {Wokaun}},\
  }\href {\doibase 10.1063/1.442690} {\bibfield  {journal} {\bibinfo  {journal}
  {J. Chem. Phys.}\ }\textbf {\bibinfo {volume} {76}},\ \bibinfo {pages} {751}
  (\bibinfo {year} {1982})}\BibitemShut {NoStop}%
\bibitem [{\citenamefont {Klingsporn}\ \emph {et~al.}(2014)\citenamefont
  {Klingsporn}, \citenamefont {Sonntag}, \citenamefont {Seideman},\ and\
  \citenamefont {{Van Duyne}}}]{klingsporn2014}%
  \BibitemOpen
  \bibfield  {author} {\bibinfo {author} {\bibfnamefont {J.~M.}\ \bibnamefont
  {Klingsporn}}, \bibinfo {author} {\bibfnamefont {M.~D.}\ \bibnamefont
  {Sonntag}}, \bibinfo {author} {\bibfnamefont {T.}~\bibnamefont {Seideman}}, \
  and\ \bibinfo {author} {\bibfnamefont {R.~P.}\ \bibnamefont {{Van Duyne}}},\
  }\href {\doibase 10.1021/jz4024404} {\bibfield  {journal} {\bibinfo
  {journal} {J. Phys. Chem. Lett.}\ }\textbf {\bibinfo {volume} {5}},\ \bibinfo
  {pages} {106} (\bibinfo {year} {2014})}\BibitemShut {NoStop}%
\bibitem [{\citenamefont {Wang}\ \emph
  {et~al.}(2007{\natexlab{a}})\citenamefont {Wang}, \citenamefont {Liu},
  \citenamefont {Zhuang}, \citenamefont {Zhang}, \citenamefont {Wang},
  \citenamefont {Xie}, \citenamefont {Wu}, \citenamefont {Ren},\ and\
  \citenamefont {Tian}}]{wang2007}%
  \BibitemOpen
  \bibfield  {author} {\bibinfo {author} {\bibfnamefont {X.}~\bibnamefont
  {Wang}}, \bibinfo {author} {\bibfnamefont {Z.}~\bibnamefont {Liu}}, \bibinfo
  {author} {\bibfnamefont {M.-D.}\ \bibnamefont {Zhuang}}, \bibinfo {author}
  {\bibfnamefont {H.-M.}\ \bibnamefont {Zhang}}, \bibinfo {author}
  {\bibfnamefont {X.}~\bibnamefont {Wang}}, \bibinfo {author} {\bibfnamefont
  {Z.-X.}\ \bibnamefont {Xie}}, \bibinfo {author} {\bibfnamefont {D.-Y.}\
  \bibnamefont {Wu}}, \bibinfo {author} {\bibfnamefont {B.}~\bibnamefont
  {Ren}}, \ and\ \bibinfo {author} {\bibfnamefont {Z.-Q.}\ \bibnamefont
  {Tian}},\ }\href {\doibase 10.1063/1.2776860} {\bibfield  {journal} {\bibinfo
   {journal} {Appl. Phys. Lett.}\ }\textbf {\bibinfo {volume} {91}},\ \bibinfo
  {pages} {101105} (\bibinfo {year} {2007}{\natexlab{a}})}\BibitemShut
  {NoStop}%
\bibitem [{\citenamefont {Lopes}\ \emph {et~al.}(2013)\citenamefont {Lopes},
  \citenamefont {Toury}, \citenamefont {{De La Chapelle}}, \citenamefont
  {Bonaccorso},\ and\ \citenamefont {Gucciardi}}]{lopes2013}%
  \BibitemOpen
  \bibfield  {author} {\bibinfo {author} {\bibfnamefont {M.}~\bibnamefont
  {Lopes}}, \bibinfo {author} {\bibfnamefont {T.}~\bibnamefont {Toury}},
  \bibinfo {author} {\bibfnamefont {M.~L.}\ \bibnamefont {{De La Chapelle}}},
  \bibinfo {author} {\bibfnamefont {F.}~\bibnamefont {Bonaccorso}}, \ and\
  \bibinfo {author} {\bibfnamefont {P.~G.}\ \bibnamefont {Gucciardi}},\ }\href
  {\doibase 10.1063/1.4812365} {\bibfield  {journal} {\bibinfo  {journal} {Rev.
  Sci. Instrum.}\ }\textbf {\bibinfo {volume} {84}},\ \bibinfo {pages} {073702}
  (\bibinfo {year} {2013})}\BibitemShut {NoStop}%
\bibitem [{\citenamefont {Kharintsev}\ \emph {et~al.}(2011)\citenamefont
  {Kharintsev}, \citenamefont {Noskov}, \citenamefont {Hoffmann},\ and\
  \citenamefont {Loos}}]{kharintsev2011}%
  \BibitemOpen
  \bibfield  {author} {\bibinfo {author} {\bibfnamefont {S.~S.}\ \bibnamefont
  {Kharintsev}}, \bibinfo {author} {\bibfnamefont {A.~I.}\ \bibnamefont
  {Noskov}}, \bibinfo {author} {\bibfnamefont {G.~G.}\ \bibnamefont
  {Hoffmann}}, \ and\ \bibinfo {author} {\bibfnamefont {J.}~\bibnamefont
  {Loos}},\ }\href {\doibase 10.1088/0957-4484/22/2/025202} {\bibfield
  {journal} {\bibinfo  {journal} {Nanotechnology}\ }\textbf {\bibinfo {volume}
  {22}},\ \bibinfo {pages} {025202} (\bibinfo {year} {2011})}\BibitemShut
  {NoStop}%
\bibitem [{\citenamefont {Downes}, \citenamefont {Salter},\ and\ \citenamefont
  {Elfick}(2006)}]{downes2006}%
  \BibitemOpen
  \bibfield  {author} {\bibinfo {author} {\bibfnamefont {A.}~\bibnamefont
  {Downes}}, \bibinfo {author} {\bibfnamefont {D.}~\bibnamefont {Salter}}, \
  and\ \bibinfo {author} {\bibfnamefont {A.}~\bibnamefont {Elfick}},\ }\href
  {https://www.osapublishing.org/DirectPDFAccess/131D9500-B918-31B6-B25DA37D8CFAF8C9{\_}117889/oe-14-23-11324.pdf?da=1{\&}id=117889{\&}seq=0{\&}mobile=no}
  {\bibfield  {journal} {\bibinfo  {journal} {Opt. Express}\ }\textbf {\bibinfo
  {volume} {14}} (\bibinfo {year} {2006})}\BibitemShut {NoStop}%
\bibitem [{\citenamefont {Roth}\ \emph {et~al.}(2006)\citenamefont {Roth},
  \citenamefont {Panoiu}, \citenamefont {Adams}, \citenamefont {Osgood},
  \citenamefont {Neacsu},\ and\ \citenamefont {Raschke}}]{roth2005}%
  \BibitemOpen
  \bibfield  {author} {\bibinfo {author} {\bibfnamefont {R.~M.}\ \bibnamefont
  {Roth}}, \bibinfo {author} {\bibfnamefont {N.~C.}\ \bibnamefont {Panoiu}},
  \bibinfo {author} {\bibfnamefont {M.~M.}\ \bibnamefont {Adams}}, \bibinfo
  {author} {\bibfnamefont {R.~M.}\ \bibnamefont {Osgood}}, \bibinfo {author}
  {\bibfnamefont {C.~C.}\ \bibnamefont {Neacsu}}, \ and\ \bibinfo {author}
  {\bibfnamefont {M.~B.}\ \bibnamefont {Raschke}},\ }\href {\doibase
  10.1364/OE.14.002921} {\bibfield  {journal} {\bibinfo  {journal} {Opt.
  Express}\ }\textbf {\bibinfo {volume} {14}},\ \bibinfo {pages} {2921}
  (\bibinfo {year} {2006})}\BibitemShut {NoStop}%
\bibitem [{\citenamefont {Issa}\ and\ \citenamefont
  {Guckenberger}(2007)}]{Issa2007}%
  \BibitemOpen
  \bibfield  {author} {\bibinfo {author} {\bibfnamefont {N.~A.}\ \bibnamefont
  {Issa}}\ and\ \bibinfo {author} {\bibfnamefont {R.}~\bibnamefont
  {Guckenberger}},\ }\href {\doibase 10.1364/OE.15.012131} {\bibfield
  {journal} {\bibinfo  {journal} {Opt. Express}\ }\textbf {\bibinfo {volume}
  {15}},\ \bibinfo {pages} {12131} (\bibinfo {year} {2007})}\BibitemShut
  {NoStop}%
\bibitem [{\citenamefont {Stockman}(2004)}]{Stockman2004}%
  \BibitemOpen
  \bibfield  {author} {\bibinfo {author} {\bibfnamefont {M.}~\bibnamefont
  {Stockman}},\ }\href {\doibase 10.1103/PhysRevLett.93.137404} {\bibfield
  {journal} {\bibinfo  {journal} {Phys. Rev. Lett.}\ }\textbf {\bibinfo
  {volume} {93}},\ \bibinfo {pages} {137404} (\bibinfo {year}
  {2004})}\BibitemShut {NoStop}%
\bibitem [{\citenamefont {Kryder}\ \emph {et~al.}(2008)\citenamefont {Kryder},
  \citenamefont {Gage}, \citenamefont {Mcdaniel}, \citenamefont {Challener},
  \citenamefont {Rottmayer}, \citenamefont {Ju}, \citenamefont {Hsia},\ and\
  \citenamefont {Erden}}]{Kryder2008}%
  \BibitemOpen
  \bibfield  {author} {\bibinfo {author} {\bibfnamefont {M.~H.}\ \bibnamefont
  {Kryder}}, \bibinfo {author} {\bibfnamefont {E.~C.}\ \bibnamefont {Gage}},
  \bibinfo {author} {\bibfnamefont {T.~W.}\ \bibnamefont {Mcdaniel}}, \bibinfo
  {author} {\bibfnamefont {W.~A.}\ \bibnamefont {Challener}}, \bibinfo {author}
  {\bibfnamefont {R.~E.}\ \bibnamefont {Rottmayer}}, \bibinfo {author}
  {\bibfnamefont {G.}~\bibnamefont {Ju}}, \bibinfo {author} {\bibfnamefont
  {Y.~T.}\ \bibnamefont {Hsia}}, \ and\ \bibinfo {author} {\bibfnamefont
  {M.~F.}\ \bibnamefont {Erden}},\ }\href {\doibase 10.1109/JPROC.2008.2004315}
  {\bibfield  {journal} {\bibinfo  {journal} {Proc. IEEE}\ }\textbf {\bibinfo
  {volume} {96}},\ \bibinfo {pages} {1810} (\bibinfo {year}
  {2008})}\BibitemShut {NoStop}%
\bibitem [{\citenamefont {Raether}(1988)}]{Raether1988}%
  \BibitemOpen
  \bibfield  {author} {\bibinfo {author} {\bibfnamefont {H.}~\bibnamefont
  {Raether}},\ }\href {\doibase 10.1007/BFb0048317} {\bibfield  {journal}
  {\bibinfo  {journal} {Springer Tracts Mod. Phys.}\ }\textbf {\bibinfo
  {volume} {111}},\ \bibinfo {pages} {136} (\bibinfo {year}
  {1988})}\BibitemShut {NoStop}%
\bibitem [{\citenamefont {Hessel}\ and\ \citenamefont
  {Oliner}(1965)}]{Hessel1965}%
  \BibitemOpen
  \bibfield  {author} {\bibinfo {author} {\bibfnamefont {A.}~\bibnamefont
  {Hessel}}\ and\ \bibinfo {author} {\bibfnamefont {A.~A.}\ \bibnamefont
  {Oliner}},\ }\href {\doibase 10.1364/AO.4.001275} {\bibfield  {journal}
  {\bibinfo  {journal} {Appl. Opt.}\ }\textbf {\bibinfo {volume} {4}},\
  \bibinfo {pages} {1275} (\bibinfo {year} {1965})}\BibitemShut {NoStop}%
\bibitem [{\citenamefont {Tikui{\vaccent{s}}is}\ \emph
  {et~al.}(2017)\citenamefont {Tikui{\vaccent{s}}is}, \citenamefont {Beran},
  \citenamefont {Cejpek}, \citenamefont {Uhl{\'{i}}{\vaccent{r}}ov{\'{a}}},
  \citenamefont {Hamrle}, \citenamefont {Va{\vaccent{n}}atka}, \citenamefont
  {Urb{\'{a}}nek},\ and\ \citenamefont {Veis}}]{Tikuisis2017}%
  \BibitemOpen
  \bibfield  {author} {\bibinfo {author} {\bibfnamefont {K.~K.}\ \bibnamefont
  {Tikui{\vaccent{s}}is}}, \bibinfo {author} {\bibfnamefont {L.}~\bibnamefont
  {Beran}}, \bibinfo {author} {\bibfnamefont {P.}~\bibnamefont {Cejpek}},
  \bibinfo {author} {\bibfnamefont {K.}~\bibnamefont
  {Uhl{\'{i}}{\vaccent{r}}ov{\'{a}}}}, \bibinfo {author} {\bibfnamefont
  {J.}~\bibnamefont {Hamrle}}, \bibinfo {author} {\bibfnamefont
  {M.}~\bibnamefont {Va{\vaccent{n}}atka}}, \bibinfo {author} {\bibfnamefont
  {M.}~\bibnamefont {Urb{\'{a}}nek}}, \ and\ \bibinfo {author} {\bibfnamefont
  {M.}~\bibnamefont {Veis}},\ }\href {\doibase 10.1016/j.matdes.2016.10.036}
  {\bibfield  {journal} {\bibinfo  {journal} {Mater. Des.}\ }\textbf {\bibinfo
  {volume} {114}},\ \bibinfo {pages} {31} (\bibinfo {year} {2017})}\BibitemShut
  {NoStop}%
\bibitem [{\citenamefont {Butt}, \citenamefont {Cappella},\ and\ \citenamefont
  {Kappl}(2005)}]{Butt2005}%
  \BibitemOpen
  \bibfield  {author} {\bibinfo {author} {\bibfnamefont {H.-J.}\ \bibnamefont
  {Butt}}, \bibinfo {author} {\bibfnamefont {B.}~\bibnamefont {Cappella}}, \
  and\ \bibinfo {author} {\bibfnamefont {M.}~\bibnamefont {Kappl}},\ }\href
  {\doibase 10.1016/j.surfrep.2005.08.003} {\bibfield  {journal} {\bibinfo
  {journal} {Surf. Sci. Rep.}\ }\textbf {\bibinfo {volume} {59}},\ \bibinfo
  {pages} {1} (\bibinfo {year} {2005})}\BibitemShut {NoStop}%
\bibitem [{\citenamefont {Hamaker}(1937)}]{Hamaker1937}%
  \BibitemOpen
  \bibfield  {author} {\bibinfo {author} {\bibfnamefont {H.}~\bibnamefont
  {Hamaker}},\ }\href {\doibase 10.1016/S0031-8914(37)80203-7} {\bibfield
  {journal} {\bibinfo  {journal} {Physica}\ }\textbf {\bibinfo {volume} {4}},\
  \bibinfo {pages} {1058} (\bibinfo {year} {1937})}\BibitemShut {NoStop}%
\bibitem [{\citenamefont {Zhang}\ \emph {et~al.}(2009)\citenamefont {Zhang},
  \citenamefont {Zhao}, \citenamefont {Huang}, \citenamefont {Yang},\ and\
  \citenamefont {Zhang}}]{Zhang2009}%
  \BibitemOpen
  \bibfield  {author} {\bibinfo {author} {\bibfnamefont {Y.-Z.}\ \bibnamefont
  {Zhang}}, \bibinfo {author} {\bibfnamefont {B.}~\bibnamefont {Zhao}},
  \bibinfo {author} {\bibfnamefont {G.-Y.}\ \bibnamefont {Huang}}, \bibinfo
  {author} {\bibfnamefont {Z.}~\bibnamefont {Yang}}, \ and\ \bibinfo {author}
  {\bibfnamefont {Y.-F.}\ \bibnamefont {Zhang}},\ }\href {\doibase
  10.1007/s11671-009-9335-5} {\bibfield  {journal} {\bibinfo  {journal}
  {Nanoscale Res. Lett.}\ }\textbf {\bibinfo {volume} {4}},\ \bibinfo {pages}
  {850} (\bibinfo {year} {2009})}\BibitemShut {NoStop}%
\bibitem [{\citenamefont {Kittel}\ \emph {et~al.}(2005)\citenamefont {Kittel},
  \citenamefont {M{\"{u}}ller-Hirsch}, \citenamefont {Parisi}, \citenamefont
  {Biehs}, \citenamefont {Reddig},\ and\ \citenamefont
  {Holthaus}}]{Kittel2005}%
  \BibitemOpen
  \bibfield  {author} {\bibinfo {author} {\bibfnamefont {A.}~\bibnamefont
  {Kittel}}, \bibinfo {author} {\bibfnamefont {W.}~\bibnamefont
  {M{\"{u}}ller-Hirsch}}, \bibinfo {author} {\bibfnamefont {J.}~\bibnamefont
  {Parisi}}, \bibinfo {author} {\bibfnamefont {S.-A.}\ \bibnamefont {Biehs}},
  \bibinfo {author} {\bibfnamefont {D.}~\bibnamefont {Reddig}}, \ and\ \bibinfo
  {author} {\bibfnamefont {M.}~\bibnamefont {Holthaus}},\ }\href {\doibase
  10.1103/PhysRevLett.95.224301} {\bibfield  {journal} {\bibinfo  {journal}
  {Phys. Rev. Lett.}\ }\textbf {\bibinfo {volume} {95}},\ \bibinfo {pages}
  {224301} (\bibinfo {year} {2005})}\BibitemShut {NoStop}%
\bibitem [{\citenamefont {Wang}\ \emph
  {et~al.}(2007{\natexlab{b}})\citenamefont {Wang}, \citenamefont {Xu},
  \citenamefont {Shimono}, \citenamefont {Tanaka},\ and\ \citenamefont
  {Yamazaki}}]{2-Wang2007}%
  \BibitemOpen
  \bibfield  {author} {\bibinfo {author} {\bibfnamefont {H.}~\bibnamefont
  {Wang}}, \bibinfo {author} {\bibfnamefont {Y.}~\bibnamefont {Xu}}, \bibinfo
  {author} {\bibfnamefont {M.}~\bibnamefont {Shimono}}, \bibinfo {author}
  {\bibfnamefont {Y.}~\bibnamefont {Tanaka}}, \ and\ \bibinfo {author}
  {\bibfnamefont {M.}~\bibnamefont {Yamazaki}},\ }\href {\doibase
  10.2320/matertrans.MAW200717} {\bibfield  {journal} {\bibinfo  {journal}
  {Mater. Trans.}\ }\textbf {\bibinfo {volume} {48}},\ \bibinfo {pages} {2349}
  (\bibinfo {year} {2007}{\natexlab{b}})}\BibitemShut {NoStop}%
\bibitem [{\citenamefont {Nix}\ and\ \citenamefont {MacNair}(1941)}]{Nix1941}%
  \BibitemOpen
  \bibfield  {author} {\bibinfo {author} {\bibfnamefont {F.~C.}\ \bibnamefont
  {Nix}}\ and\ \bibinfo {author} {\bibfnamefont {D.}~\bibnamefont {MacNair}},\
  }\href {\doibase 10.1103/PhysRev.60.597} {\bibfield  {journal} {\bibinfo
  {journal} {Phys. Rev.}\ }\textbf {\bibinfo {volume} {60}},\ \bibinfo {pages}
  {597} (\bibinfo {year} {1941})}\BibitemShut {NoStop}%
\bibitem [{\citenamefont {Staelin}, \citenamefont {Morgenthaler},\ and\
  \citenamefont {Kong}(1998)}]{Staelin1998}%
  \BibitemOpen
  \bibfield  {author} {\bibinfo {author} {\bibfnamefont {D.}~\bibnamefont
  {Staelin}}, \bibinfo {author} {\bibfnamefont {A.~W.}\ \bibnamefont
  {Morgenthaler}}, \ and\ \bibinfo {author} {\bibfnamefont {J.~A.}\
  \bibnamefont {Kong}},\ }\href@noop {} {\emph {\bibinfo {title}
  {Electromagnetic waves}}}\ (\bibinfo  {publisher} {Prentice Hall},\ \bibinfo
  {year} {1998})\BibitemShut {NoStop}%
\end{thebibliography}%

\end{document}